# Higher-Spin Theories - Part II: enter dimension three

**Gustavo Lucena Gómez**[*]
*Physique Théorique et Mathématique & International Solvay Instituts*
*Université Libre de Bruxelles, Campus Plaine C.P. 231, B-1050 Bruxelles, Belgium*
*E-mail:* glucenag@ulb.ac.be

These notes aim at providing a pedagogical and pedestrian introduction to the subject and assume no previous knowledge apart from that of general relativity. We shall first recall the "frame" formulation of the later theory, then particularize it to three dimensions, and will end those preliminaries by reviewing the formulation of three-dimensional gravity as a gauge theory governed by a Chern–Simons action. An analogous path is then followed for higher-spin fields at the free level. Once the equivalent Chern–Simons action is established thereof, it is then explained how one can formulate three-dimensional higher-spin theories *at the non-linear level* by considering higher-spin Lie algebras. We then move on to commenting on what has already been done in the context of these theories and what interesting areas of research are currently under investigation.

---

[*]Speaker.

**Introduction**

**Disclaimer:** these notes are meant to be pedagogical and self-contained. However, as the goal is to lead the reader in a rather straight course to an interacting picture for higher-spins in three-dimensions, some computations shall be left as exercises, and comments as well as references may not be made or given in an exhaustive fashion. This being said, the only required knowledge to read the present notes is that of standard general relativity, with the exception of some generic knowledge about free higher-spin fields in the metric-like formulation — the latter being part of R. Rahman's lectures on higher-spin theories in dimension four and greater [1]. Similarly, the rest of this introduction has been freed of the standard contextualization of the field of higher-spins in general — also found in [1] — and mostly focuses on the three-dimensional peculiarities.

Any kind of comment regarding these notes would be highly appreciated.[1]

$$\therefore \quad \therefore \quad \therefore$$

Over the years, the study of three-dimensional gravity has proved most fruitful (see [2] for a review). Briefly put, the 1986 discovery of the Brown–Henneaux central charge for pure gravity with negative cosmological constant [3] and subsequent investigations proved the field to be an incredibly rich laboratory for quantum gravity. Moreover, the reformulation of it as a Chern–Simons gauge theory [4, 5] has made its study easier and even more appealing. Let us recall that three-dimensional gravity is topological, which can be seen as the key feature making it less complicated than its higher-dimensional cousins. This is also one of the primary reasons for the study of three-dimensional higher spins; the theory is still topological and hence simpler than the corresponding higher-dimensional setups. However, as our theory including higher spins now presumably has a much richer CFT dual (for we are adding boundary degrees of freedom to it), its study has an interest of its own. Furthermore, the study of higher-spin black holes, which is notoriously intricate in any dimension higher than three, also contributes to making the field an exciting one to investigate. Finally, let us note that a key result in the field is that of Blencowe in 1989 [6], which is essentially the analogue of [4, 5] for pure gravity; namely, three-dimensional higher-spins, much like pure gravity in that dimension, admit a Chern–Simons gauge formulation. However, in the case of higher spins the result is doubly important: first, analogously to the spin-2 case, it makes the handling of the subject quite easier and is more suited to the study of some of its aspects (as for example asymptotic symmetries), but, more importantly and this has no analogue in general relativity, it allows one to introduce couplings in a rather straightforward way — which in the case of higher-spins is intricate even in dimension three, not to mention torturing in higher dimensions. Such a formulation of three-dimensional higher-spins is actually much used in the related literature nowadays, and we take it as the point to be reached by these lectures.

---

[1]In arXiv version 2 and beyond corrections have been made upon the first version, published in *PoS*.

Section 1 hereafter shall be concerned with recalling first the frame (or vielbein) formulation of gravity and second the Chern–Simons (or gauge) formulation thereof, in three dimensions. Higher-spins are therefore only introduced in Section 2, where the same logic is followed; the generalized frame formulation is first exposed, at the free level, and we then move on to the Chern–Simons formulation of the theory, where interactions are rather easy to introduce by means of finite- or infinite-dimensional higher-spin algebras. In Section 3 we end these lectures by briefly commenting on some recent developments in the field.

Let us emphasize once again that these notes contain nothing new, except maybe the ambition of being pedestrian as well as somewhat pedagogical in spirit.

## 1. Gravity as a gauge theory in 3D

What shall be done in the next section for higher-spin fields is here reviewed in the context of pure gravity. We shall first recall, in any dimension, the formulation of gravity in terms of the vielbein and the spin-connection [7] and will then specialize to three-dimensions, where this formalism helps making contact with the gauge formulation of gravity as a Chern–Simons gauge theory.

Let us stress already that, in the case of gravity, all of the above reformulations can be carried out at the non-linear (interacting) level, whereas for higher-spins we will only introduce interactions once the Chern–Simons formulation is at hand (see next section). Note, however, that it is in principle possible to introduce interactions at the level of the metric-like formulation for higher-spins, but it is far from being as clean as doing it in the gauge picture.

### 1.1 The frame formulation of gravity

This subsection relies, among others, on [7], which we recommend to the reader unfamiliar with the subject, for only a quick review is provided in the present notes. Many other references on this subject are available, among which we shall highlight the mathematically oriented one [8] as well as the earlier one [9]. Note that only the three-dimensional version of the present subsection is useful for us but, as it is not much effort to do so, for the sake of completeness we shall start in general dimension $D$ and will only particularize to $D=3$ at the end of the subsection.

#### 1.1.1 The vielbein

Pure gravity is described by the Einstein–Hilbert action in $D$ spacetime dimensions with cosmological constant $\Lambda$ (here for $c=1$, that is assumed through these notes unless otherwise specified):

$$S_{\text{EH}} \equiv S_{\text{EH}}[g] \equiv \frac{1}{16\pi \text{G}} \int_{\mathcal{M}_D} (R - 2\Lambda)\sqrt{-g}\, \mathrm{d}^D x, \tag{1.1}$$

where G is the $D$-dimensional Newton constant, $g$ is the determinant of the metric $g_{\mu\nu}$, $R$ is the Ricci scalar and $\mathcal{M}_D$ is the spacetime manifold. The equations of motion one derives from the above action read

$$G_{\mu\nu} + \Lambda g_{\mu\nu} = 0, \tag{1.2}$$

where $G_{\mu\nu} \equiv R_{\mu\nu} - \frac{1}{2}g_{\mu\nu}R$ is the Einstein tensor and $R \equiv g^{\mu\nu}R_{\mu\nu}$ is the Ricci scalar, the contraction of the Ricci tensor with the inverse metric $g^{\mu\nu}$. The usual rewriting of the above equations without involving the Ricci scalar is then

$$R_{\mu\nu} = g_{\mu\nu}\Lambda\frac{2}{D-2}. \tag{1.3}$$

Let us now introduce the so-called vielbein by the relation

$$g_{\mu\nu} \equiv e_\mu^a e_\nu^b \eta_{ab}, \tag{1.4}$$

with our conventions for the signature of the Minkowski metric $\eta_{ab}$ being $(-+\cdots+)$. The Latin indices are usually referred to as "frame" indices. The relation is, however, invariant under the so-called local Lorentz transformations (LLTs) of the vielbein

$$e_\mu^a(x) \to \Lambda^a_{\ b}(x) e_\mu^b(x), \tag{1.5}$$

with the matrix $\Lambda(x) \in \mathrm{SO}(D-1,1)$ (the Lorentz group). Now the vielbein is a $D \times D$ matrix at each spacetime point, of which we can eliminate as many components as the dimension of the Lie algebra so$(D-1,1)$ (at each spacetime point), that is, $D(D-1)/2$, which leaves us with $D(D+1)/2$ independent components: the number of independent components of the $D$-dimensional metric. Thinking of (1.4) as a mere change of variables for gravity, the transformation law (1.5) simply originates in that the change of variables is not one-to-one and some redundancy is introduced (which we just saw can be "gauged away" using LLTs).

**Remark** : albeit dubbed "local Lorentz transformations", let us point out that those are quite different from the usual Lorentz transformations. Indeed, the later act on spacetime indices and are rigid (or global) transformations, that is, the Lorentz matrices do not depend on spacetime coordinates. The local Lorentz transformations are quite different in that respect, as they do not act on spacetime indices but on frame indices and, furthermore, the Lorentz matrices thereof do depend on spacetime coordinates. The situation can be understood as having a Lorentz invariance freedom at every spacetime point for our frame indices. This symmetry is thus to be thought of as a gauge symmetry of the formulation to come.

Our vielbein is a hybrid object; it has both a spacetime and a frame index. We already displayed its transformation rules with respect to its frame index (LLTs), which resulted from a redundancy in our change of variables. With respect to its spacetime index, the tensor nature of the metric forces it to transform as a covector under the diffeomorphism group.

The spacetime indices are always raised and lowered using the metric $g_{\mu\nu}$ and its inverse, but, the metric governing the frame indices is always the Minkowski one $\eta_{ab}$. This is actually related to a conceptually important fact: the so-called tangent frame defined by the vielbein is orthogonal at any spacetime point, as the following relations illustrate

$$e_\mu^a e_b^\mu = \delta_b^a, \quad e_\mu^a e_a^\nu = \delta_\mu^\nu, \tag{1.6}$$

where the Latin indices have been raised and lowered with $\eta_{ab}$. The vielbein $e^\mu_a$, with both indices swapped, is called the inverse vielbein (and it is indeed so if we think of it as a matrix). Also note the useful relations

$$\eta_{ab} = g_{\mu\nu} e^\mu_a e^\nu_b, \quad \sqrt{-g} = e \equiv \det(e^a_\mu). \tag{1.7}$$

We thus see that, at every spacetime point, the vielbein is providing us with some flat frame in which the metric looks flat — the tangent frame. For a more in-depth understanding of the geometrical interpretation of this object we refer to [8].

**Remark** : note that the coordinate system itself is also some tangent frame, but it is not orthogonal at every spacetime point. Indeed, the principle of equivalence only states that at any spacetime point one can find a *locally* inertial rest frame, but, with respect to that frame, the rest of the spacetime does not generally look flat (unless the geometry is flat of course). That is, in the metric formalism, when moving away from the spacetime point where we made the geometry look locally flat, we keep the same frame but the metric changes and is no longer trivial. It is the precise opposite that happens in the vielbein formulation: all the changes in the geometry are encoded in the vielbein, which changes when we move away from some spacetime point so to remain an orthogonal frame, in which the metric looks trivial ($\eta_{ab}$) at every spacetime point.

### 1.1.2 The spin-connection

Working in the tangent frame will force us to consider various objects having frame indices, such as the vielbein, that we already encountered, and we want to be able to derive such objects. For simplicity, let us first focus on a frame vector, that is, an object having only one frame index (upstairs), $V^a$. Let us recall how we proceed in gravity when we want to derive objects having spacetime indices: we start with $\partial_\mu V^\nu$ (for, say, a spacetime vector), and imposing that it is again a tensor under the diffeomorphism group we are uniquely[2] led to introducing a spacetime connection $\Gamma^\mu_{\nu\rho}$ with specific transformation rules (not those of a tensor) such that $\nabla_\mu V^\nu \equiv \partial_\mu V^\nu + \Gamma^\rho_{\mu\nu} V^\nu$ is a covariant object (transforms as a tensor). Recall the transformation rules for the spacetime connection:

$$\Gamma'^\rho_{\mu\nu}(x') \equiv \partial_\sigma x'^\rho \left( \partial'^2_{\mu\nu} x^\sigma + \partial'_\mu x^\alpha \partial'_\nu x^\beta \Gamma^\sigma_{\alpha\beta} \right). \tag{1.8}$$

Now, proceeding in an analogous way and requiring the derivative of our tangent frame vector to be a *tangent frame* tensor, we are again uniquely led to introducing the so-called spin-connection $\omega^{ab}_\mu$; a hybrid object having the following transformation rules under local Lorentz transformations (avoiding to spell out the spacetime indices, which remain unaffected by such transformations):

$$\omega'^a{}_b(x) = (\Lambda^{-1})^a{}_c(x) d\Lambda^c{}_b(x) + (\Lambda^{-1})^a{}_c(x) \omega^c{}_d(x) \Lambda^d{}_b(x), \tag{1.9}$$

which ensure that the tangent frame covariant derivative,

$$D_\mu V^a = \partial_\mu V^a + \omega^a{}_{\mu b} V^b,$$

---

[2]We mean unique in the sense that the transformation rules (1.8) are uniquely dictated by requiring $\nabla_\mu V^\nu$ to be covariant.

is an "LLT tensor" (transforms covariantly under LLTs). The resemblance between the above expression and the law (1.8) for the spacetime connection is worth noticing. Let us again insist on that, in the above transformation rule, spacetime indices are not affected, and hence nor are the coordinates.

Some comments are now in order. First, the above derivation rules — for both spacetime and tangent frame indices — are of course more complicated when one derives higher-order tensors (see [7]). Second, both rules are to be combined when deriving hybrid objects (such as the vielbein or the spin-connection). That "full" derivation will be noted $\hat{\nabla}$. This will help us distinguish all three types of derivation rules one could use to derive a hybrid object: one could derive it with respect to its spacetime indices ($\nabla$), with respect to its tangent frame indices ($D$), or with respect to both ($\hat{\nabla}$). As an example, we give

$$\hat{\nabla}_\mu e^a_\nu = \partial_\mu e^a_\nu + \omega^a{}_{\mu b} e^b_\nu - \Gamma^\rho_{\mu\nu} e^a_\rho. \tag{1.10}$$

We now want to solve for the spin-connection, that is, to impose conditions such that we can uniquely find some $\omega = \omega[e]$. In the metric formalism, one imposes metric-compatibility as well as zero-torsion for the spacetime connection, which uniquely leads to the well-known Christoffel expression for $\Gamma^\mu_{\nu\rho}$ in terms of the metric and its derivatives. A similar thing happens in the tangent frame. Indeed, imposing metric-compatibility as well as zero-torsion one uniquely finds

$$\omega^{ab}_\mu[e] = 2e^{\nu[a} \partial_{[\mu} e^{b]}_{\nu]} - e^{\nu[a} e^{b]\sigma} e_{\mu c} \partial_\nu e^c_\sigma. \tag{1.11}$$

At this point the reader might complain about the fact that we did not display the tangent frame torsion. The later definition is actually very simple. First recall that, in the standard approach, torsion can be expressed as $T^\mu = \nabla \mathrm{d}x^\mu$ (with some indices left out for conciseness). We are thus tempted to try $T^a \equiv De^a$ (with again two spacetime indices left out), which can indeed be seen to yield the correct notion of torsion (the tangent frame analogue of what we know in spacetime). An analogous intuition is also valid for the notion of metric-compatibility, which in spacetime means $\nabla_\mu g_{\nu\rho} = 0$ and translates to the tangent frame as $D\eta = 0$. Note, however, that the latter is simply equivalent to $\omega^{ab}_\mu = -\omega^{ba}_\mu$. It is interesting to notice that, in spacetime, it is the zero-torsion condition which is equivalent to the symmetry of the connection (in its two lower indices), whereas in the frame it is the metric-compatibility condition which implies antisymmetry in the two Latin indices. Although we are not giving all the details of how to arrive at (1.11) (see [7]), the key point here is that one *can* solve for $\omega = \omega[e]$ and, furthermore, the conditions uniquely leading to the solution are precisely the "tangent-frame translation" of the conditions one usually imposes in standard gravity. This analogy between the frame picture and the usual metric formulation can actually be pushed further, which we do in the sequel.

### 1.1.3 The curvature

Being aware of the aforementioned analogies between the spacetime and the spin-connection, and recalling the known expression for the Riemann tensor $R[\Gamma]$, one is naturally led to consider the following object, also commonly called the curvature:

$$R^{ab}[\omega] \equiv \mathrm{d}\omega^{ab} + \omega^a{}_c \wedge \omega^{cb}. \tag{1.12}$$

It is a conceptually important point that we are also led to such an expression if we simply notice that the spin-connection transforms under LLTs just like a standard Yang–Mills gauge potential, and indeed, the above expression is precisely the standard Yang–Mills field strength for $\omega$. It is therefore a hybrid object we are dealing with, and we further note that it has two spacetime indices and two tangent frame ones, and that it is a tensor with respect to both types of indices (under diffeomorphisms and LLTs respectively). This "dual nature" goes even further, for the above tensor is blessed with two Bianchi-like identities:

$$\begin{aligned} R[\omega]^{ab} \wedge e_b &= R_{\mu\nu\rho}{}^a + R_{\nu\rho\mu}{}^a + R_{\rho\mu\nu}{}^a = 0, \\ \mathrm{d}R[\omega]^{ab} + \omega^a{}_c \wedge R[\omega]^{cb} - R[\omega]^{ac} \wedge \omega_c{}^b &= D_\mu R_{\nu\rho}{}^{ab} + D_\nu R_{\rho\mu}{}^{ab} + D_\rho R_{\mu\nu}{}^{ab} = 0, \end{aligned} \quad (1.13)$$

the first one being "purely gravitational" (with no analogue in Yang–Mills theory), and the second one being the standard gauge-theory identity simply stemming from the definition of the field strength. These Bianchi identities are heavily reminiscent of the ones endowing the Riemann curvature tensor. Actually, it is one of the most important basic results of the frame formulation of gravity that the following relation holds:

$$R[\Gamma]^\rho_{\mu\nu\,\sigma} = R[\omega]_{\mu\nu ab} e^{a\rho} e^b_\sigma. \quad (1.14)$$

Note that, knowing the relations linking the spacetime and spin-connection to the metric and vielbein respectively, a direct check of such a relation seems doable but incredibly tedious. However, there exists a trick, which is the mere evaluation of $[\nabla_\mu, \nabla_\nu] e^\rho_a = 0$ and which leads to the result more easily (it is left as an exercise for the reader).

Another analogy between both formulations of the curvature is their transformation under a variation of the spacetime and spin-connection respectively. Those read

$$\begin{aligned} \delta R[\Gamma]^\rho_{\mu\nu\,\sigma} &= \nabla_\mu(\delta\Gamma^\rho_{\nu\sigma}) - \nabla_\nu(\delta\Gamma^\rho_{\mu\sigma}), \\ \delta R[\omega]_{\mu\nu ab} &= D_\mu(\delta\omega_{\nu ab}) - D_\nu(\delta\omega_{\mu ab}), \end{aligned} \quad (1.15)$$

the latter of which will be useful when going back from the so-called first-order formalism to the 1.5-order formalism, that we have been dealing with so far without giving it that name (see next subsection).

Last but not least we point out maybe the most compelling reason to think of both the above curvatures as being the analogue of one another, that is, they both appear when one considers the commutator of two covariant derivatives ($D$ and $\nabla$ respectively), which is really the object characterizing the curvature of spacetime, that is, how much it fails to be flat — and we again refer to [7].

### 1.1.4 The action

We are finally ready to rewrite the Einstein–Hilbert action (1.1) in terms of the vielbein. Indeed, the relation (1.14) leaves us only with the problem of rewriting the infinitesimal spacetime volume element, but, fortunately, the following relations are easily derived:

$$e\mathrm{d}^D x = e^0 \wedge \cdots \wedge e^{D-1} = \frac{1}{D!} \epsilon_{a_1 \ldots a_D} e^{a_1} \wedge \cdots \wedge e^{a_D} = \frac{e}{D!} \epsilon_{\mu_1 \ldots \mu_D} \mathrm{d}x^{\mu_1} \wedge \cdots \wedge \mathrm{d}x^{\mu_D}, \quad (1.16)$$

our conventions being $\epsilon_{0...D-1} \equiv 1$. Indeed, plugging the above rewriting of the infinitesimal volume element $dV$ as well as relation (1.14) in the action (1.1) and further using

$$e^{a_1} \wedge \cdots \wedge e^{a_p} \wedge e^{b_1} \wedge \cdots \wedge e^{b_q} = -\epsilon^{a_1...a_p b_1...b_q} dV, \quad (1.17)$$

we finally find, for the $\Lambda = 0$ case,

$$S_{\text{EH}}[g[e]] = \frac{1}{(D-2)!16\pi G} \int_{\mathcal{M}_D} \epsilon_{abc_1...c_{D-2}} e^{c_1} \wedge \cdots \wedge e^{c_{D-2}} \wedge R[\omega[e]]^{ab}$$
$$\equiv S_{1.5}[e,\omega[e]] \equiv S_{1.5}[e], \quad (1.18)$$

which in three-dimensions reads (now including the obvious contribution from the cosmological constant)

$$S_{1.5}[e] = \frac{1}{16\pi G} \int_{\mathcal{M}_3} \epsilon_{abc} e^a \wedge (R^{bc}[\omega] - \frac{\Lambda}{3} e^b \wedge e^c), \quad (1.19)$$

and where "1.5" stands for "1.5-order" formalism, meaning that the spin-connection is thought of as depending on the vielbein, so that the latter obeys a second order differential equation, and thus we are indeed somewhat in between the second-order formalism (Einstein–Hilbert) and the first-order one, to be introduced hereafter.[3] For the three-dimensional epsilon symbol we also use the convention $\epsilon_{123} \equiv 1$. Note that the dependence of the spin-connection on the vielbein is usually dropped when writing the action, just as we did for the last expression above. It is rather clear that, because the actions are equivalent, finding the equations of motion for the vielbein that the above action yields and going back to the metric formulation one should find the Einstein equations. This is a nice exercise that we shall not comment on more.

A natural thing to do now is to consider the same action, but forgetting about the relation between the spin-connection and the vielbein, that is, both objects are treated independently. Then, the variational principle yields equations for $\omega$ as well, in addition to those for the vielbein. As it turns out, these "equations of motion" for $\omega$ are precisely the zero-torsion condition. Therefore, if we further demand that $\omega$ be antisymmetric in its two frame indices (which we might very well do and still consider it independent of the vielbein), its equation of motion allows us to solve for it, obtaining the expression (1.11). This way of thinking about the frame formulation of the action is called "first-order" formalism, because now the spin-connection is to be thought of as an auxiliary field (for which we can solve), and before integrating it out the vielbein thus obeys a first-order differential equation (which gives back Einstein equations if we plug in it the functional dependence of the spin-connection on the vielbein). To stress that it depends on $e$ and $\omega$ independently and to distinguish it from the 1.5-order action (1.19), the first order action will be noted $S_{\text{FO}}[e,\omega]$ (but looks exactly like (1.19) except for the fact that $\omega$ is no longer to be understood *off-shell* as a functional of the vielbein and its derivatives).

---

[3]For the 1.5-order formalism, which is now standard material found for example in [7], we point out that previous to the standard references [10, 11] there was usage of it in [12].

Note that, in order to find the equations of motion for the spin-connection, one first uses the aforegiven formula (1.15). Then, combining the vielbein postulate [7] with the integration formula

$$\int \mathrm{d}^D x \sqrt{-g}\nabla_\mu V^\mu = \int \mathrm{d}^D x \partial_\mu(\sqrt{-g}V^\mu) + \int \mathrm{d}^D x \sqrt{-g}T^\nu_{\nu\mu}V^\mu, \tag{1.20}$$

one obtains that some combination of the torsion is equal to zero; an equation that one has to hit with some vielbein in order to finally get the zero-torsion condition. Such a derivation of the zero-torsion condition is recommended for the inexperienced reader.

From now on and until we reach Section 2 we shall work in three dimensions, where we can perform the standard "dual" rewriting

$$R[\omega]_a \equiv \frac{1}{2}\epsilon_{abc}R[\omega]^{bc} \quad \Leftrightarrow \quad R[\omega]^{ab} = -\epsilon^{abc}R[\omega]_c, \tag{1.21}$$

and do the same for $\omega^a$ itself, thus obtaining:

$$R[\omega]_a = \mathrm{d}\omega_a + \frac{1}{2}\epsilon_{abc}\omega^b \wedge \omega^c. \tag{1.22}$$

The action (1.19) at $\Lambda = 0$ can thus be rewritten as

$$S_{\mathrm{FO}}[e,\omega] = \frac{2}{16\pi \mathrm{G}}\int_{\mathcal{M}_3} e^a \wedge R[\omega]_a, \tag{1.23}$$

which we recall can only be written in three dimensions. Note that we have moved to the first-order formalism, for it is the one we shall start from in order to pass to the Chern–Simons formulation.

Before moving to the next subsection, let us display the linearized equations of motion corresponding to the first-order formalism. The reason why we only need display the linearized equations of motion is that, in the next section, we shall start from a linearized higher-spin theory in order to try to build its non-linear completion. The linearized higher-spin equations of motion will thus be expressed in the frame formalism and, in order to have something to compare them to, we give hereafter the linearized equations of motion in the frame formalism for the $s = 2$ case. In linearizing the vielbein, we have adopted the notation $e \rightarrow \bar{e} + v$, where $\bar{e}$ is the background dreibein associated with the background metric (here that of three-dimensional anti-de Sitter spacetime) via the usual formula (1.4). As for the spin-connection, we have used $\omega \rightarrow \bar{\omega} + \omega$, where $\bar{\omega}$ is some background fixed value, that on-shell would be related to $\bar{e}$ via the usual zero-torsion condition — that is, via equation (1.11). We find these excitations to satisfy

$$\begin{aligned}\mathrm{d}\omega^a + \epsilon^{abc}\bar{\omega}_b \wedge \omega_c + \Lambda\epsilon^{abc}\bar{e}_b \wedge v_c &= 0,\\ \mathrm{d}v^a + \epsilon^{abc}\bar{\omega}_b \wedge v_c + \epsilon^{abc}\bar{e}_b \wedge \omega_c &= 0,\end{aligned} \tag{1.24}$$

which can be rewritten as

$$\begin{aligned}D\omega^a + \Lambda\epsilon^{abc}\bar{e}_b \wedge v_c &= 0,\\ Dv^a + \epsilon^{abc}\bar{e}_b \wedge \omega_c &= 0,\end{aligned} \tag{1.25}$$

where the first one above is the linearized zero-torsion condition for the metric-compatible spin-connection $\omega^a$ and the second one is simply the linearized equation of motion for the dreibein.

One may check that the above equations are invariant under linearized diffeomorphisms as well as infinitesimal local Lorentz transformations, as they should be.

### 1.2 3D gravity as a Chern–Simons theory

As has been argued in the previous subsection, three-dimensional pure gravity is equivalent to the first-order formalism, with the vielbein and the spin-connection being independent variables (the latter being an auxiliary field). Starting from the latter formulation, we shall now discuss the result of Achúcarro–Townsend–Witten [4, 5] in which yet another formulation of gravity is found (in three-dimensions), namely that of a gauge theory described by a Chern–Simons action with a connection one-form $A_\mu$ taking values in the Lie algebra of isometries of the vacuum solution. The said algebra being either Minkowski, anti-de Sitter or de Sitter, the relevant Lie algebras underlying our yet-to-be-formulated gauge description of 3D gravity will be respectively iso(2,1), so(2,2) or so(3,1).

The way in which we shall proceed is backwards, that is, we will start from the gauge theory we claim to be equivalent to three-dimensional gravity and will then show it to be so. Even though this is not completely standard material, we shall be rather pragmatic in spirit and refer to [5] for a more detailed discussion.

#### 1.2.1 Intuitive and handwaving "bla bla"

The frame formulation of gravity has made many physicists try to combine the vielbein and the spin-connection into some iso$(D-1,1)$-valued one-form gauge field. Indeed, the vielbein (resp. the spin-connection) looks like an appealing candidate for the role of the coefficient of the gauge connection $A$ corresponding to the translation generators (resp. Lorentz generators) of iso$(D-1,1)$. This is so firstly because the dimensions match and, furthermore because, as mentioned in the previous subsection, when formulated in terms of the vielbein and spin-connection gravity already has some of the taste of a Yang–Mills-like gauge theory. However, there is an easy intuitive reason why this is probably doomed to fail in dimension four (or at the very least somewhat unnatural). Indeed, looking back at (1.18) for $D=4$ we see that it is of the schematic form (for $\Lambda=0$):

$$S \sim \int_{\mathcal{M}_4} e \wedge e \wedge (\mathrm{d}\omega + \omega \wedge \omega), \qquad (1.26)$$

so that the corresponding Yang–Mills-like action should look somewhat like

$$S \sim \int_{\mathcal{M}_4} \mathrm{Tr}(A \wedge A \wedge (\mathrm{d}A + A \wedge A)), \qquad (1.27)$$

which does not exist in gauge theory (because the trace of the wedge product of two algebra-valued one-forms is identically zero). However, in $D=3$, we have the well-known Chern–Simons action, which roughly looks like

$$S \sim \int_{\mathcal{M}_4} \mathrm{Tr}(A \wedge (\mathrm{d}A + A \wedge A)); \qquad (1.28)$$

precisely what one feels like trying when looking at (1.19) for $\Lambda = 0$ in dimension three!

The other indication that the three-dimensional scenario is specially suited for establishing such a correspondence has to do with bilinear forms on the relevant Lie algebras. Indeed, if one is to build a Chern–Simons action (or any Yang–Mills-like action for that matter) for some Lie algebra $\mathcal{G}$, one should first make sure that there exists some invariant, non-degenerate, symmetric and bilinear form on it. In the $\Lambda = 0$ case, it turns out that iso$(D-1,1)$ admits such a form only for $D = 3 \ldots$!

### 1.2.2 The gauge algebras

Let us start by considering the gauge algebras that we will have to work with in the sequel, which will also allow us to fix the conventions thereof. From now on we stick to $D = 3$, for which the commutation relations of our three different gauge algebras can be packaged into

$$[J_a, J_b] = \epsilon_{abc}J^c, \qquad [J_a, P_b] = \epsilon_{abc}P^c, \qquad [P_a, P_b] = -\lambda\epsilon_{abc}J^c, \tag{1.29}$$

where the Latin indices $a, b, c = 1, 2, 3$ are raised and lowered with the three-dimensional Minkowski metric[4] $\eta_{ab}$ and its inverse $\eta^{ab}$, which are also chosen to have signature $(-++)$. Note that we have used the "three-dimensional rewriting"

$$J_a \equiv \frac{1}{2}\epsilon_{abc}J^{bc} \quad \Leftrightarrow \quad J^{ab} \equiv -\epsilon^{abc}J_c, \tag{1.30}$$

where $J_{ab}$ are the usual Lorentz generators ($P_a$ are of course the translation ones). For $\lambda = 0$, $\lambda < 0$ and $\lambda > 0$, the above relations describe respectively iso$(2,1)$, so$(2,2)$ and so$(3,1)$.

As aforementioned, iso$(D-1,1)$ admits a non-degenerate and invariant (symmetric and real) bilinear form[5] only for $D = 3$, which is unique in the space of such forms.[6] It reads

$$(J_a, P_b) = \eta_{ab}, \qquad (J_a, J_b) = (P_a, P_b) = 0. \tag{1.31}$$

As for so$(D-1,2)$ and so$(D,1)$, they admit a non-degenerate, invariant bilinear form for any $D$, the particularization of which to $D = 3$ reads

$$(J_a, J_b) = \eta_{ab}, \qquad (J_a, P_b) = 0, \qquad (P_a, P_b) = -\lambda\eta_{ab}. \tag{1.32}$$

For $D \neq 3$ they are both simple and the higher-dimensional equivalent of the above form is therefore unique (up to normalization) and proportional to the Killing form. For $D = 3$, however, the AdS one becomes semi-simple and undergo the splittings

$$\text{so}(2,2) \simeq \text{sl}(2|\mathbb{R}) \oplus \text{sl}(2|\mathbb{R}), \tag{1.33}$$

so that, in addition to the above form it also admits (1.31).

---

[4]This is important since if one takes the indices to be euclidean the commutation relations would describe, e.g. for $\lambda = 0$, iso(3) instead of iso(2,1).

[5]For a general treatment of Lie algebras we recommend for example [13].

[6]Note that it is not the Killing form, the latter being degenerate because iso(2,1) is not semi-simple.

**Remark** : note that (1.32) is degenerate for $\lambda = 0$, which is the reason iso$(D-1,1)$ only admits a non-degenerate form for $D=3$, (1.31), which is a specificity of the three-dimensional case, as can be seen by noting that it corresponds to the invariant $\epsilon_{abc}J^{ab}P^c$, which can only be constructed in three-dimensions. It is thus the "epsilon magic" which is really at work here.

In the $\Lambda \geq 0$ case we do not have any choice for the bilinear form to use, but, as for the $\Lambda < 0$ case we have the freedom of choosing our bilinear form among the two above ones. However, it seems somewhat more natural to use also in that case the one which also endows the isometry algebra of three-dimensional Minkowski spacetime. This choice will not be further justified in the present work (except by the fact that it will lead to some Chern–Simons action which will indeed reproduce the Einstein–Hilbert one), and we refer the interested reader to [5] for more discussions on the subject.

As we shall be most interested in the AdS case, let us already point out that the splitting (1.33) of so(2,2) explicitly reads

$$[J_a^+, J_b^+] = \epsilon_{abc}J^{+c}, \qquad [J_a^-, J_b^-] = \epsilon_{abc}J^{-c}, \qquad [J_a^+, J_b^-] = 0, \tag{1.34}$$

where

$$J_a^\pm \equiv \frac{1}{2}\left(J_a \pm \frac{1}{\sqrt{-\lambda}}P_a\right). \tag{1.35}$$

Before moving to the next subsection, we also note that, when expressed in terms of the $J^\pm$ generators of so(2,2), the above form (1.35) reads

$$(J_a^+, J_b^+) = \frac{1}{2}\eta_{ab}, \qquad (J_a^-, J_b^-) = -\frac{1}{2}\eta_{ab}, \qquad (J_a^+, J_b^-) = 0, \tag{1.36}$$

which we recall corresponds to the $\Lambda < 0$ case.

### 1.2.3 The action

Now that the algebraic aspects have been dealt with, let us work out the equivalence at the level of the actions between some Chern–Simons term with connection one-form $A_\mu$ living in one of the above three-dimensional isometry algebras and three-dimensional Einstein–Hilbert gravity with corresponding cosmological constant.

We begin by proving the equivalence in the $\Lambda = 0$ case. The identification of the degrees of freedom is the following:

$$A_\mu \equiv e_\mu^a P_a + \omega_\mu^a J_a, \tag{1.37}$$

where the generators $J_a, P_a$ of iso(2,1) satisfy (1.29) at $\lambda = 0$. If we now plug this expansion into the Chern–Simons action term below and use (1.31) for the scalar product (trace) a straightforward computation yields

$$\begin{aligned} S_{\text{CS}}[A] &\equiv \kappa \int_{\mathcal{M}_3} \text{Tr}\left(A \wedge dA + \frac{2}{3} A \wedge A \wedge A\right) \\ &= \kappa \int_{\mathcal{M}_3} e^a \wedge R[\omega]_a \\ &\equiv \kappa 16\pi G S_{\text{FO}}[e,\omega], \end{aligned} \tag{1.38}$$

where we have used

$$\epsilon_{abc}\epsilon^{ade} = \delta_b^e \delta_c^d - \delta_b^d \delta_c^e. \tag{1.39}$$

We thus conclude that, upon our identification (1.37), $S_{\text{CS}}[A]$ equals $S_{\text{FO}}[e,\omega]$ provided we set $\kappa = 1/16\pi G$, which is called the Chern–Simons level. We also note that the matching of the actions assumes that the vielbein is invertible. We shall not dwell on this interesting issue here, and refer to [5] for an interesting discussion..

Let us point out that there is another, perhaps more elegant way of showing the equivalence, which is based on the identity

$$\text{d}\,\text{Tr}\left(A \wedge \text{d}A + \frac{2}{3} A \wedge A \wedge A\right) = \text{Tr}\left((\text{d}A + A \wedge A) \wedge (\text{d}A + A \wedge A)\right), \tag{1.40}$$

for, as we know, regardless of the algebra the Trace of any even "wedge-power" of $A$ vanishes identically. In other words,

$$\text{d}\,\text{Tr}\left(A \wedge \text{d}A + \frac{2}{3} A \wedge A \wedge A\right) = \text{Tr}\left(F[A] \wedge F[A]\right), \tag{1.41}$$

where we have used the definition $F[A] \equiv \text{d}A + A \wedge A$. This implies that, for a four-dimensional manifold $\mathcal{M}_4$ such that $\text{d}\mathcal{M}_4 = \mathcal{M}_3$, the "four-dimensional" action

$$S'_{\text{CS}} \equiv S'_{\text{CS}}[A] \equiv \kappa \int_{\mathcal{M}_4} \text{Tr}\left(F[A] \wedge F[A]\right) \tag{1.42}$$

is equal to the above $S_{\text{CS}}[A]$ (1.38). The final step is then to realize that it is possible to identify some components of $A$ with the dreibein and the spin-connection in a way that makes the above action equal to $S_{\text{FO}}[e,\omega]$ (upon reducing the four-dimensional integral to the three-dimensional one on the boundary $\mathcal{M}_3$ of course). This is actually the way it is proved in the original paper [5]. However, one does not gain much time compared with the above demonstration and moreover this detour not only contains a few subtleties but also requires considering four-dimensional action terms which might render the discussion a little confusing, which is why we have chosen to follow a more straightforward procedure here.

Before moving to the next subsection, let us work out — for we shall need them — the equations of motion derived from the above Chern–Simons action. In terms of the gauge connection they read $R[A] = 0$, as is well known. In terms of $e$ and $\omega$ we easily find the corresponding expressions:

$$\begin{aligned} D_\mu e_\nu^a - D_\nu e_\mu^a &= 0, \\ \partial_\mu \omega_\nu^a - \partial_\nu \omega_\mu^a + \epsilon^{abc} \omega_{b\mu} \omega_{c\nu} &= 0, \end{aligned} \tag{1.43}$$

where we use the standard abuse of notation

$$D_\mu e_\nu^a \equiv (D_\mu e_\nu)|_{P_a}, \tag{1.44}$$

where $|_{P_a}$ means taking the component along the $P_a$ generators. Note that, as it should be, the above equations of motion do coincide, at the linearized level, with the $\Lambda = 0$ version of (1.24).

**Remark** : note that the pairing between $J_a$ and $P_a$ in the bilinear form (1.31) violates dimensionality, since the generator $P_a$ has inverse length dimension, while $J_a$ and $\eta_{ab}$ are dimensionless. This happens because we choose to isolate the Newton constant G in front of the Chern–Simons action (1.38), where $\kappa = 1/16\pi G$. With a trace (bilinear form) in agreement with the dimensionality of the generators, the Chern–Simons action with $\kappa = 1/16\pi G$ would actually fail to be dimensionless. Strictly speaking, the correct thing to do would be to remove the Newton constant from $\kappa$ and rescale (1.31) as $(J_a, P_b) = \eta_{ab}/G$, which is then consistent since G has the dimensions of length for $D = 3$. This would also make the action manifestly dimensionless.[7]

### 1.2.4 The gauge transformations

There is a last non-trivial check to do before one can safely claim the two theories to be equivalent; namely, we need verify the gauge transformations on both sides to be the same. Indeed, while both the first-order formulation and the Chern–Simons action are manifestly invariant under diffeomorphisms, in the first-order formulation we also have the local Lorentz transformations as gauge symmetries, whereas in the Chern–Simons picture we have the full iso(2,1) gauge symmetries. As we shall now show, the homogeneous part of the iso(2,1) gauge symmetries are easily seen to correspond to the LLTs in the first-order formalism but, as for the infinitesimal gauge translations of iso(2,1), one has to show that they are not extra gauge symmetries (which would be bad for our rewriting of the action would then eliminate degrees of freedom in some sense) but, rather, that they correspond to some combination of the symmetries of the first-order formalism action.

Now the gauge transformations in the Chern–Simons picture are parametrized by a zero-form gauge parameter taking values in the gauge algebra,

$$u \equiv \rho^a P_a + \tau^a J_a, \tag{1.45}$$

with $\rho^a$ and $\tau^a$ being infinitesimal parameters, and the transformation law for the gauge connection (sitting in the adjoint representation of the gauge algebra) is $A \to A + \delta A$ with

$$\delta A_\mu = \partial_\mu u + [A_\mu, u]. \tag{1.46}$$

Upon now plugging the expression for $u$ and the decomposition of $A$ in terms of the dreibein and spin-connection in the above equation we can read off the variations of $e$ and $\omega$, which read

$$\begin{aligned} \delta e_\mu^a &= \partial_\mu \rho^a + \epsilon^{abc} e_{b\mu} \tau_c + \epsilon^{abc} \omega_{b\mu} \rho_c, \\ \delta \omega_\mu^a &= \partial_\mu \tau^a + \epsilon^{abc} \omega_{b\mu} \tau_c. \end{aligned} \tag{1.47}$$

These are all the (infinitesimal) local symmetries of the action in the gauge (Chern–Simons) picture and they can be decomposed into those generated by $\rho^a$,

$$\begin{aligned} \delta e_\mu^a &= \partial_\mu \rho^a + \epsilon^{abc} \omega_{b\mu} \rho_c, \\ \delta \omega_\mu^a &= 0, \end{aligned} \tag{1.48}$$

---

[7]We thank P. Kessel for pointing out this issue.

and those generated by $\tau^a$,
$$\begin{aligned} \delta e_\mu^a &= \epsilon^{abc} e_{b\mu} \tau_c, \\ \delta \omega_\mu^a &= \partial_\mu \tau^a + \epsilon^{abc} \omega_{b\mu} \tau_c. \end{aligned} \quad (1.49)$$

Moreover, as we already explained, the Chern–Simons term is also manifestly invariant under diffeomorphisms because it is written in terms of forms. The diffeomorphisms act by the well-known formula
$$\delta A_\mu = \xi^\nu \partial_\nu A_\mu + A_\nu \partial_\mu \xi^\nu, \quad (1.50)$$
where $\xi^\nu$ is some infinitesimal vector. We may again use the identification (1.37) to now simply find
$$\begin{aligned} \delta e_\mu^a &= \xi^\nu \partial_\nu e_\mu^a + e_\nu^a \partial_\mu \xi^\nu = \xi^\nu (\partial_\nu e_\mu^a - \partial_\mu e_\nu^a) + \partial_\mu (e_\nu^a \xi^\nu), \\ \delta \omega_\mu^a &= \xi^\nu \partial_\nu \omega_\mu^a + \omega_\nu^a \partial_\mu \xi^\nu = \xi^\nu (\partial_\nu \omega_\mu^a - \partial_\mu \omega_\nu^a) + \partial_\mu (\omega_\nu^a \xi^\nu), \end{aligned} \quad (1.51)$$
which are the usual diffeomorphism transformations in the frame picture. Then, when dealing with the first-order action we also have the local Lorentz transformations, which act on $e$ and $\omega$ as in (1.5) and (1.9) respectively, the infinitesimal version of which is
$$\begin{aligned} \delta e_\mu^a &= -\alpha^a{}_b e_\mu^b, \\ \delta \omega_\mu^a &= -\alpha^a{}_b \omega_\mu^b + \frac{1}{2} \epsilon^{abc} \partial_\mu \alpha_{bc}, \end{aligned} \quad (1.52)$$
as is easily derived taking $\alpha^a{}_b$ to be an element of the Lorentz algebra with coefficients depending on spacetime coordinates. These are the local symmetries of the frame formulation.

Now, the later LLTs are quite easily seen to be in one-to-one correspondence with the gauge transformations generated by the parameters $\tau^a$ on the Chern–Simons side. Indeed, upon setting
$$\alpha^{ab} = -\epsilon^{abc} \tau_c \quad \Leftrightarrow \quad \tau^a = \frac{1}{2} \epsilon^{abc} \alpha_{bc}, \quad (1.53)$$
one sees that (1.52) and (1.49) agree with one another. As for the infinitesimal gauge transformations of the gauge picture generated by the $\rho^a$'s the story is a little more subtle, and indeed at first sight one wonders what they could correspond to in the frame formulation. Actually, we shall need use both the equations of motion and the invariance under local Lorentz transformations to make them match with the infinitesimal diffeomorphisms or, differently put, we will show that the gauge transformations generated by the $\rho^a$'s are somehow not extra gauge transformations but, rather, on-shell they are simply some combination of diffeomorphisms and LLTs. Let us then try to relate the infinitesimal parameter $\xi^\mu$ to $\rho^a$. We are tempted to try
$$\rho^a = \xi^\mu e_\mu^a, \quad (1.54)$$
which yields
$$\begin{aligned} (\delta_\xi - \delta_\rho) e_\mu^a &= \xi^\nu (\partial_\nu e_\mu^a - \partial_\mu e_\nu^a) + \epsilon^{abc} \xi^\nu e_{b\nu} \omega_{c\mu}, \\ (\delta_\xi - \delta_\rho) \omega_\mu^a &= \xi^\nu (\partial_\nu \omega_\mu^a - \partial_\mu \omega_\nu^a) + \partial_\mu (\omega_\nu^a \xi^\nu). \end{aligned} \quad (1.55)$$
Now, the first terms in the right hand sides of the two above equations are seen to be the "abelian" part of the equations of motion (1.43), so we try making these terms exactly the whole equations of motion, which yields

$$\begin{aligned}(\delta_\xi - \delta_\rho)e^a_\mu &= \xi^\nu(D_\nu e^a_\mu - D_\mu e^a_\nu) + \epsilon^{abc}\xi^\nu e_{b\mu}\omega_{c\nu},\\(\delta_\xi - \delta_\rho)\omega^a_\mu &= \xi^\nu(\partial_\nu \omega^a_\mu - \partial_\mu \omega^a_\nu + \epsilon^{abc}\omega_{b\nu}\omega_{c\mu}) + \xi^\nu \epsilon^{abc}\omega_{b\mu}\omega_{c\nu} + \partial_\mu(\omega^a_\nu\xi^\nu).\end{aligned} \quad (1.56)$$

We now see that the first terms are proportional to the equations of motion whereas the last terms are local Lorentz transformations with parameter

$$\alpha^{ab} = -\epsilon^{abc}\xi^\nu\omega_{c\nu} \quad \Leftrightarrow \quad \tau^a = \xi^\nu\omega^a_\nu, \qquad (1.57)$$

so that gauge transformations generated by $\rho^a$, which are also named infinitesimal gauge translations, indeed correspond to diffeomorphisms in the frame formulation (up to LLTs and EoMs). As already stated, this is well, since the point was to check that there are no extra gauge symmetries. Let us also stress that this fact is truly a three-dimensional feature and does not happen in dimension four and greater. Actually, this is precisely what prevents one from writing gravity in dimension four and greater as a gauge theory simply by gauging the isometry group of the vacuum and employing a gauge-connection valued therein. We might thus roughly say that "only in three dimensions is gravity a true gauge theory", and even there, we see that its action is that of Chern–Simons, which is not of the Yang–Mills type that we are more used to in standard gauge theory. Note that a Yang–Mills action term in three dimensions would propagate scalar degrees of freedom, unlike gravity which propagates none in dimension three.

This achieves the proof of the equivalence for the $\lambda = 0 = \Lambda$ case. Three-dimensional gravity is thus a gauge theory for the gauge group iso(2,1), the Poincaré group (for zero cosmological constant).

**Remark** : in dimensions greater than three one may also want to interpret gravity as a gauge theory of some type for the corresponding gauge group. However, as we just said in dimension four and greater the first-order action (1.23) is only invariant under the homogeneous Poincaré group so(2,1), not under the whole of iso(2,1) — that is, gauge translations do not leave the action invariant. The action (1.23) is of course invariant under coordinate reparametrization (diffeomorphisms), but those Lie derivatives do not correspond (in the first order formalism at least) to gauge translations. Only in three-dimensions does that miracle happen, thus allowing us to rewrite three-dimensional gravity as a gauge theory for the whole Poincaré group. More information can be found in [7, 9] and we shall not further comment on that point.

### 1.2.5 Cosmological constant and chiral copies

As announced, what we shall be interested in is the case with negative cosmological constant. Following the same reasoning as in the $\lambda = 0$ case (with same identifications of $e$ and $\omega$ via (1.37) and same bilinear form on the algebra) the equivalence can again be proven between the frame formulation and the Chern–Simons one based on so(2,2). As this calculation is really close to the one we have just performed we shall not do it again and we just quote what is different in the Chern–Simons picture, that is, the gauge

transformations now read

$$\begin{aligned}\delta e_\mu^a &= \partial_\mu \rho^a + \epsilon^{abc} e_{b\mu} \tau_c + \epsilon^{abc} \omega_{b\mu} \rho_c, \\ \delta \omega_\mu^a &= \partial_\mu \tau^a + \epsilon^{abc} \omega_{b\mu} \tau_c - \lambda \epsilon^{abc} e_{b\mu} \rho_c, \end{aligned} \tag{1.58}$$

and the way in which they correspond to diffeomorphisms and LLTs is the same as before.

The identification of the actions now requires one to set

$$\lambda = \Lambda, \tag{1.59}$$

so that the parameter $\lambda$ appearing in (1.29) is precisely the cosmological constant $\Lambda$, and $l$ is the AdS radius defined in the usual way by $\Lambda \equiv -1/l^2$.

Now, because of the splitting of so(2,2) into two chiral copies of sl(2|$\mathbb{R}$) (1.33), it is possible to rewrite the Chern–Simons action term for $\Gamma_\mu \in$ so(2,2) as the sum of two Chern–Simons actions, each of them having their connections $A$ and $\tilde{A}$ in the first and second chiral copy of sl(2|$\mathbb{R}$) respectively. In the sequel we shall only deal with the first chiral copy but we wanted to still call the connection thereof $A$, which is why we have changed notations at this point. Actually, the decomposition of $\Gamma$ in terms of $e$ and $\omega$ is quite helpful in formulating this splitting precisely, for we see that

$$\Gamma = e^a P_a + \omega^a J_a = (\omega^a + \frac{e^a}{l}) J_a^+ + (\omega^a - \frac{e^a}{l}) J_a^- \equiv A^a J_a^+ + \tilde{A}^a J_a^- \equiv A + \tilde{A}, \tag{1.60}$$

where the $J_a^\pm$'s are defined by (1.35). Now, taking into acount both the commutations relations (1.34) and the bilinear form (1.36) written in terms of $J_a^\pm$, we see that the Chern–Simons action term for so(2,2) can be split as follows

$$S_{\text{CS}}[\Gamma = A + \tilde{A}] = S_{\text{CS}}[A] + \tilde{S}_{\text{CS}}[\tilde{A}]. \tag{1.61}$$

In the above equation among actions, the left-hand side has the same prefactor as the flat-space action (1.38), namely $\kappa = 1/16\pi G$. However because the chiral generators as defined in (1.35) involve $l$, the prefactor for both chiral copies in the above right-hand side needs to be $\kappa = l/16\pi G$ instead. Note that for the splitting of the kinetic piece one needs only notice that $(J_a^\pm, J_b^\mp) = 0$ whereas also $[J_a^\pm, J_b^\mp] = 0$ is needed to prove the splitting of the interaction piece. Both chiral copies $S_{\text{CS}}[A]$ and $\tilde{S}_{\text{CS}}[\tilde{A}]$ are the same actions except for one difference, which is that the $J_a^+$'s and $J_a^-$'s are equipped with bilinear forms having opposite signs, as per (1.36). This is why the second chiral term is distinguished and denoted by $\tilde{S}$. Equivalently, if one prefers to have both chiral copies equipped with the same bilinear form, one can instead declare

$$\begin{aligned} A &\equiv (\omega^a + \frac{e^a}{l}) T_a, \\ \tilde{A} &\equiv (\omega^a - \frac{e^a}{l}) T_a, \end{aligned} \tag{1.62}$$

with the $T_a$ generators of sl(2|$\mathbb{R}$) satisfying the same commutation relations and scalar products as the $J_a^+$ ones (we changed notations not to confuse the reader). The decomposition of the action then reads

$$S_{\text{CS}}[\Gamma = e/l + \omega] = S_{\text{CS}}[A] - S_{\text{CS}}[\tilde{A}]. \tag{1.63}$$

It is of course no longer true that $\Gamma = A + \tilde{A}$ (nor does it make sense to write so anymore), but this decomposition in which the connections $A$ and $\tilde{A}$ both lie in the first chiral copy of sl(2|$\mathbb{R}$) (to put it that way) is handier, as the two action functionals are the same now also when seen as functionals of the components $A^a$ and $\tilde{A}^a$ — they are truly the same action functionals now. This formulation, where we only need a single sl(2|$\mathbb{R}$), will be of much use in the sequel, where we shall only treat the first chiral copy for many of our purposes. This is also the formulation that is most often encountered in the literature.

Note that the equations of motion now read

$$F[A] = 0, \quad F[\tilde{A}] = 0, \tag{1.64}$$

which, when combined as $F[A] \pm F[\tilde{A}] = 0$ and subsequently linearized are seen to yield those written in (1.24) (in the linearized limit). Note that the gauge transformations are also split now, but we shall not display them here for the sake of simplicity.

**Remark** : unlike in the flat-space case, in the $\Lambda \neq 0$ case we could have chosen the other bilinear form (1.32) instead, respectively to a linear supperposition of both these bilinear forms. This alternative choice would lead to a plus relative sign in (1.63) instead of a minus, respectively to an arbitrary linear combination thereof. All these inequivalent actions would yield the same equations of motion. This is particularly manifest in the 'splitted' form, since the relative factor in (1.32) evidently does not affect the derivation of (1.64). The choice of bilinear form may be a delicate point. It is discussed e.g. in [5], where it is mentionned how this choice encompasses all values of $\Lambda$ in a smooth way, whereas the alternative choice of bilinear form is singular at $\Lambda = 0$.

## 2. Higher-spin gravity in 3D

What has been done for pure gravity in the previous section will now be carried out for higher-spin fields ($s > 2$). However, the formulation of interacting higher-spin fields (with themselves and with gravity) is notoriously tricky. A lot of progress has been made in that direction, including a long term effort by M. A. Vasiliev, but a complete description to all orders at the level of the action is not yet available... in $D > 3$ ! For more information on the subject we refer to R. Rahman's lectures [1].

In the metric-like or frame formulation of higher-spin fields, in dimensions greater than three, building consistent interactions is rather intricate. First of all, as noted in [14], the non-vanishing Weyl tensor in $D \geq 4$ precludes the existence of so-called "hypergravity" (a spin-$\frac{5}{2}$ field minimally coupled to gravity; in some sense the first non-trivial gravitational higher-spin interaction). Secondly, dealing with the frame formalism requires so-called "extra fields" and "extra gauge symmetry" — these are auxiliary fields and associated gauge symmetries one is forced to introduce in order to formulate free higher-spin fields in the frame formalism in dimension four and greater [15]. These facts essentially complicate the introduction of interactions, although as we know Vasiliev's equations do exist.

In $D = 3$ these two complications do not arise. Indeed, the Weyl tensor vanishes in three dimensions, which does allow for interaction terms involving the minimal coupling to gravity [14, 16]. Also, as noted for example in the first sections of [17], the frame formulation of three-dimensional (free) higher-spin fields does not require so-called "extra fields" and "extra gauge symmetry". Nevertheless, in dimension three there exists a much more practical tool to introduce consistent interactions, which is precisely the Chern–Simons formulation.

Now, as we shall describe in the next subsection, one can formulate higher-spin fields in some analogue of the frame formalism for gravity that we reviewed in Subsection 1.1 (at the free level) and then from there move on, in Subsection 2.2, to the Chern–Simons picture for higher spins — much in the spirit of what was done for pure gravity. The problem of introducing interactions will then be easily dealt with, as it will be equivalent to the purely algebraic problem of finding suitable higher-spin Lie algebras (see Subsection 2.2). The formulation one then arrives to, firstly introduced by Blencowe [6], namely a Chern–Simons gauge theory for some (finite or infinite-dimensional) gauge algebra containing so(2,2) (in the AdS case), is the basis for almost every study of three-dimensional higher-spin gravity today, and is what the present section is devoted to reach.

## 2.1 The frame formulation of free higher spins

As aforementioned, some analogue of the frame formalism for gravity also exists for higher-spin fields [15]. Let us stress that all of what we shall present in this subsection takes place at the linearized level. Although we shall be mainly interested in the $AdS_3$ case, we shall develop as much of the material as possible in arbitrary dimension and on a generic constant-curvature background spacetime. Note that, although all of the following discussion can be generalized to fermions, our focus will be on bosons, for the sole reason that it is simpler in a first approach. Part of the material exposed in this subsection is also reviewed in [18].

### 2.1.1 The metric formalism

Let us first review the Fronsdal (or metric) formulation of (free) higher-spin fields [19]. As this was covered in R. Rahman's lecture [1] we shall be rather sketchy, providing only what is needed in order to then move to the frame formulation.

The Fronsdal equations of motion for a spin-$s$ gauge field described by a rank-$s$ symmetric tensor $\varphi_{\mu_1...\mu_s}$ and propagating on the Minkowski $D$-dimensional background are given by

$$F_{\mu_1...\mu_s} \equiv \Box\varphi_{\mu_1...\mu_s} - s\partial_{(\mu_1}\partial^\lambda \varphi_{\mu_2...\mu_s)\lambda} + \frac{s(s-1)}{2}\partial_{(\mu_1}\partial_{\mu_2}\varphi_{\mu_3...\mu_s)\lambda}{}^\lambda = 0, \qquad (2.1)$$

where $F$ is the so-called Fronsdal tensor, which should be thought of as a higher-spin equivalent of the linearized Ricci tensor (which it boils down to for spin 2). Our symmetrization parenthesis have weight one, so that e.g. $2A_{(ij)} \equiv A_{ij} + A_{ji}$. The above equations are invariant under the gauge transformations

$$\delta\varphi_{\mu_1...\mu_s} = s\partial_{(\mu_1}\xi_{\mu_2...\mu_s)}, \quad \text{with } \xi^\lambda{}_{\lambda\mu_3...\mu_{s-1}} = 0. \qquad (2.2)$$

One can verify that the above equations of motion are equivalent to the ones obtained from varying the action

$$S = \int d^D x \, \varphi^{\mu_1...\mu_s}(F_{\mu_1...\mu_s} - \frac{(s-1)s}{4}\eta_{(\mu_1\mu_2}F_{\mu_3...\mu_s)\lambda}{}^\lambda), \qquad (2.3)$$

where the expression in parenthesis is the higher-spin analogue of the linearized Einstein tensor. Note that the gauge invariance of the above Lagrangian uses the double-trace constraint, and hence at the level of the action we need to impose that constraint "by hand", as it does not generically hold off-shell. In fact, the gauge invariance of the above Lagrangian requires the double-trace constraint $\varphi^{\lambda\rho}{}_{\lambda\rho\mu_5...\mu_s} = 0$ to hold. As one can check, this condition can be obtained by considering derivatives of the Fronsdal tensor F (which is on-shell zero), so that on-shell it is automatically imposed, as for example at the level of theequations of motion. However, at the level of the action (off-shell) we need to impose this constraint separately, and one can verify that it is preserved by the above gauge symmetries. The above action is thus the higher-spin analogue of what we would obtain if we were to linearize the Einstein–Hilbert action (1.1) (in brackets we find the analogue of the linearized Einstein tensor), and the above equations of motion (2.1) are the higher-spin counterpart of the linearized version of (1.3) (at $\Lambda = 0$ of course).

Let us now move to fields propagating on constant-curvature backgrounds. We are thus looking for equations that should now be invariant under

$$\delta\varphi_{\mu_1...\mu_s} = s\nabla_{(\mu_1}\xi_{\mu_2...\mu_s)}, \quad \text{with } \xi^\lambda{}_{\lambda\mu_3...\mu_{s-1}} = 0, \qquad (2.4)$$

where $\nabla$ stands for the covariant derivative (see previous section) associated with the background metric $\bar{g}_{\mu\nu}$ (that we will choose to be anti-de Sitter later on). The equations of motion are now

$$\begin{aligned}\hat{F}_{\mu_1...\mu_s} &\equiv F_{\mu_1...\mu_s} + \Lambda\Big(((s^2+(D-6)s-2(D-3))\varphi_{\mu_1...\mu_s} \\ &+ s(s-1)\bar{g}_{(\mu_1\mu_2}\varphi_{\mu_3...\mu_s)\lambda}{}^\lambda\Big) \\ &= 0,\end{aligned} \qquad (2.5)$$

where $\hat{F}$ is the "AdS Fronsdal tensor" and the Fronsdal tensor $F$ itself is now understood as in (2.1) but with all derivatives replaced with covariant derivatives with respect to the background metric. Again imposing the double trace constraint on our field the free Lagrangian is fixed by the requirement of gauge invariance and reads exactly as (2.3) but with $F$ replaced by $\hat{F}$. The analogies with (1.1) and (1.3) are again quite clear.

It is quite important to note that the above equations of motion and Lagrangians are fixed by the requirement of invariance under the corresponding gauge transformations. As explained in R. Rahman's lectures [1], the interactions are "even more" constrained and although Vasiliev's equations are fully non-linear, they still lack a satisfactory corresponding action principle.

**Remark** : in three spacetime dimensions, the usual notion of spin for massless particles (the helicity) reduces to a mere distinction between bosons and fermions, that is, the little group is trivial. Nonetheless, one may wish to consider the same four-dimensional free equations describing some tensor field but in dimension three. It is then easy to see that, apart from the scalar (dual to the spin 1) and the spin-1/2 field, no degrees of freedom can propagate [20]. In particular, higher-spins do not propagate any local degree of freedom in dimension three, and neither does a graviton or a Rarita–Schwinger field. However, one may still wish to call a fully symmetric rank-s tensor satisfying the $D = 3$-projected Fronsdal equation a spin-s field. That is, of course, what we mean by a higher spin in three dimensions. In [6], Blencowe obtained precisely that object (or, rather, its translation in terms of the generalized dreibein and spin-connection) by means of projecting directly the four-dimensional equations written in terms of the frame fields onto three dimensions.

### 2.1.2 The vielbein and spin-connection

Let us now try to formulate the above higher-spin free kinematics along the lines of the frame formulation of gravity. However, the reformulation of gravity in terms of the vielbein and spin-connection was carried out at the non-linear level, whereas here we only have a linear theory to start from (see comments at the end of the previous subsection) so that we shall remain at the linearized level. Therefore, instead of the relation (1.4), what we are trying to generalize to the higher-spin case, rather, is its linearized version

$$\varphi_{\mu\nu} = 2\bar{e}^a_{(\mu} v_{\nu)a}, \tag{2.6}$$

which is simply derived by plugging $g_{\mu\nu} \equiv \bar{g}_{\mu\nu} + \varphi_{\mu\nu}$ in (1.4) and defining $\bar{e}$ to be the background vielbein, associated with $\bar{g}_{\mu\nu}$, and defined together with $v^a_\mu$ by $e \equiv \bar{e} + v$. The above relation is now invariant under

$$\delta v^a_\mu = \alpha^a{}_b \bar{e}^b_\mu, \tag{2.7}$$

for $\alpha^a{}_b \in \mathrm{so}(D-1,1)$ (remember that Latin indices are raised and lowered with $\eta_{ab}$).

The above change of variables is then generalizable to higher-spin fields. Indeed, let us introduce some generalized vielbein $e^{a_1 \ldots a_{s-1}}_\mu$. Of course, we have no higher-spin analogue of the full metric at hand, so that the only thing we can do is declare this object to be its own excitation (that is, we assume that the background generalized vielbeins vanish[8]) and try to relate it to $\varphi_{\mu_1 \ldots \mu_s}$ in a sensible way that generalizes (2.6). This was done in the founding paper [15], resulting in the arbitrary-spin expression

$$\varphi_{\mu_1 \ldots \mu_s} \equiv s \bar{e}^{a_1}_{(\mu_1} \ldots \bar{e}^{a_{s-1}}_{\mu_{s-1}} e_{\mu_s)a_1 \ldots a_{s-1}}, \tag{2.8}$$

which is invariant under

$$\delta e^{a_1 \ldots a_{s-1}}_\mu = \bar{e}_{b\mu} \alpha^{b,a_1 \ldots a_{s-1}}, \tag{2.9}$$

for

$$\alpha^{(b,a_1 \ldots a_{s-1})} = 0. \tag{2.10}$$

---

[8]Because the generalized (or higher-spin) vielbeins have no background values, we shall stick to the notation $e^{a_1 \ldots a_{s-1}}_\mu$.

Note that, because Latin indices are raised and lowered with the Minkowski metric the last condition above indeed coincides, in the $s = 2$ case, with the matrix $(\alpha^a{}_b) \in so(D-1,1)$. Now, in the standard frame approach to higher spins the generalized vielbein is chosen to be an irreducible Lorentz tensor in its frame indices, that is, we choose it to be symmetric and traceless in those same indices, i.e. we impose the conditions

$$e_\mu^{a_1...a_{s-1}} = e_\mu^{(a_1...a_{s-1})}, \quad e_{\mu b}{}^{ba_1...a_{s-3}} = 0, \tag{2.11}$$

the latter of which ensures the double-trace constraint on the field $\varphi_{\mu_1...\mu_s}$, which is a way of checking that we are propagating the correct number of DoFs. Now, with such a choice of generalized vielbeins, our generalized LLT parameter $\alpha$ will have to satisfy

$$\alpha^{b,a_1...a_{s-1}} = \alpha^{b,(a_1...a_{s-1})}, \quad \alpha^{b,a_1...a_{s-3}c}{}_c = 0, \tag{2.12}$$

which, together with (2.10) implies

$$\alpha^{b,(a_1...a_{s-1})}{}_b = 0. \tag{2.13}$$

Then, much like in gravity the vielbein is just a covector with respect to its spacetime index, and in the present formulation our generalized vielbein will have covariant transformation rules under the "generalized diffeomorphisms" (2.4) such that its application to (2.8) reproduces (2.4). What we obtain is simply

$$\delta e_\mu^{a_1...a_{s-1}} = (s-1)\bar{e}_{\nu_1}^{(a_1}...\bar{e}_{\nu_{s-1}}^{a_{s-1})}\nabla_\mu \xi^{\nu_1...\nu_{s-1}}. \tag{2.14}$$

Again proceeding along the lines of what is known for gravity one introduces some generalized spin-connection $\omega_\mu^{a,b_1...b_{s-1}}$, satisfying the same conditions (2.10), (2.12) and (2.13) as the parameter $\alpha$:

$$\omega_\mu^{b,a_1...a_{s-1}} = \omega_\mu^{b,(a_1...a_{s-1})}, \quad \omega_\mu{}^{b,a_1...a_{s-3}c}{}_c = 0, \quad \omega_\mu^{(b,a_1...a_{s-1})} = 0, \tag{2.15}$$

which together imply

$$\omega_\mu{}^{b,(a_1...a_{s-2})}{}_b = 0. \tag{2.16}$$

Note that for $s = 2$ the condition $\omega_\mu^{(b,a_1...a_{s-1})} = \alpha_\mu^{(b,a_1...a_{s-1})} = 0$ is implied by the antisymmetry in the two only Latin indices then carried by $\omega$ and $\alpha$.

The last comment we shall make in the present subsection is that a "dual" rewriting analogous to (1.21) can also be performed in the arbitrary-spin case so that, ultimately, our first order formalism will deal with the generalized dreibein $e_\mu^{a_1...a_{s-1}}$ and a generalized spin-connection $\omega_\mu^{a_1...a_{s-1}}$ (in dimension three) — a rewriting one can always perform in three-dimensions.

### 2.1.3 The action and the equations of motion

In his pioneering work [15], Vasiliev identified a first-order action for the generalized vielbeins and spin-connections such that, when solving for the auxiliary field $\omega$ in terms of $e$ and further recalling the definition (2.8), one recovers an action functional coinciding with that of Fronsdal (2.3). For the sake of conciseness we only give here its four-dimensional spin-$s$ expression at $\Lambda = 0$, which reads

$$S = \int d^4 x \epsilon^{\mu\nu\rho\sigma} \epsilon^{abc}{}_\sigma \omega_{\rho,b,a}{}^{i_1...i_{s-2}} \left( \partial_\mu e_{\nu, i_1...i_{s-2}c} - \frac{1}{2} \omega_{\mu,\nu, i_1...i_{s-2}c} \right). \tag{2.17}$$

Such an action, if we believe it to be equivalent to the Fronsdal one (which it is), will be invariant under generalized LLT as well as generalized diffeomorphisms. However, as one can check, it is also invariant under an extra gauge transformation, acting only on the spin-connection [15]. That extra gauge parameter can of course be checked to vanish in the $s = 2$ case but, most importantly, in the arbitrary-spin case it also vanishes at $D = 3$! The reason why this is a key point is that one of the difficulties in formulating higher-spin theories stems from the fact that this extra gauge symmetry calls for so-called "extra (gauge) fields" associated with it (much like we can think of the spin-connection as the gauge field associated with the LLT gauge symmetry), and one is actually led through an iterative procedure which introduces several of them. Dealing with such extra gauge fields is a notorious source of inconveniences in the higher-spin context and the fact that they are not needed in three dimensions can be thought of as being one of the reasons why the three-dimensional case is simpler to deal with.

Actually, reference [15] only deals with the four-dimensional Minkowski case, and one has to refer to [21] in order to get the corresponding (A)dS expression. As for the three-dimensional scenario, it was first treated in [6], where the usual frame expressions for free higher-spin fields were projected onto three-dimensional spacetimes and then completed to yield a fully interacting theory. Before giving its expression, note that we shall not display frame-index contraction explicitly when it is thought to be obvious (see below). On the AdS$_3$ spacetime background, that we are most interested in, the obtained spin-$s$ expression is thus (now in terms of the "dualized" spin-connection):

$$S = \int e \wedge D\omega - \frac{1}{2} \epsilon^{abc} \bar{e}_a \wedge (\omega_b \wedge \omega_c + \Lambda e_b \wedge e_c), \tag{2.18}$$

and the corresponding spin-$s$ equations of motion thus read[9]

$$\begin{aligned} D\omega^{a_1...a_{s-1}} + \Lambda \epsilon^{aba_1} \bar{e}_a \wedge e_b{}^{a_2...a_{s-1}} &= 0, \\ De^{a_1...a_{s-1}} + \epsilon^{aba_1} \bar{e}_a \wedge \omega_b{}^{a_2...a_{s-1}} &= 0, \end{aligned} \tag{2.19}$$

and indeed one can verify that they enjoy no extra gauge invariance of any sort — only diffeomorphisms and local Lorentz transformations. For example, the first term in the above action evidently implies a contraction of all the indices of $e$ with all the indices of (the dualized) $\omega$, their index structure being the same. The same goes, for example, for both terms within the brackets in the action; we assume contraction of all indices except the ones that are displayed (and which are contracted with the epsilon tensor). Let us further stress that, since the spin-2 dreibeins are denoted respectively by $\bar{e}_a$ (background) and $v_a$ (excitation), there can be no confusion with some higher-spin dreibein of which we display only one frame index, as in the above action — recall that the higher-spin dreibeins and spin-connections are always assumed to have zero background values. Finally, let us point out that the background spin-connection for the spin-2 enters the action via the covariant derivative $D$.

---

[9]For $s = 2$ these equations of course boil down to those given in (1.24), only recalling that, in that case the excitation is given the letter $v_a$.

Although we don't give the proof [17] that the above action is indeed equivalent to the Fronsdal one we point out the enlightening similarity of the above equations with the linearized equations (1.24); the structure is really the same, and all we have done is deal with the extra indices in the only possible way. Let us also make it clear that the apparent discrepancy one might seem to notice between the above action and (2.17) simply lies in the fact that (2.17) is given on a flat background, where $\bar{e}$ is the trivial matrix and $\bar{\omega}$ is zero. With those precisions in mind it becomes obvious that the above action is simply a projected version of (2.17), given here at $\Lambda \neq 0$.

## 2.2 Chern–Simons action for non-linear 3D higher-spin gravity

The idea is now that, much in the spirit of what we did for pure gravity, we shall rewrite the above action (2.18) for higher spins as a Chern–Simons term whose gauge connection one-form takes values in some Lie algebra, the coefficients of which shall be identified with the generalized dreibein and spin-connection. Once again we shall proceed backwards, that is, we shall give some Chern–Simons action together with some identification of the components of its connection and then show how our action (2.18) is reproduced (at the free level).

### 2.2.1 Requirements at the linearized level

From the previous section it should be obvious that the action we are now looking for is some Chern–Simons term for a gauge connection taking values in an algebra containing $\mathrm{sl}(2|\mathbb{R})$. What is now to be investigated is what requirements are imposed on such an algebra by the matching with (2.18) at the linearized level. Note that, of course, the action we look for is the difference of two copies of the Chern–Simons action for independent combinations of the dreibein and spin-connection, like in (1.63). However, as we shall see, much of the discussion can be carried over considering only the first copy (at least the purely algebraic considerations).

Let the $T_a$'s be our spin-2 generators, the coefficients of which are associated with $e^a + \omega^a$ (and the corresponding minus sign for the other chiral copy). Now, as we have seen in the previous subsection, the higher-spin off-shell degrees of freedom[10] we need to accommodate for come in the form of the generalized dreibeins $e^{a_1 \ldots a_{s-1}}$ and spin-connections $\omega^{a_1 \ldots a_{s-1}}$, which are symmetric in their (frame) indices as well as traceless. The combination $e^{a_1 \ldots a_{s-1}} + \omega^{a_1 \ldots a_{s-1}}$ is therefore to be identified with the coefficient of some higher-spin generator $T_{a_1...a_{s-1}}$, that we may assume to be symmetric and traceless in its indices — and correspondingly for the other copy. As is easy to check, the number of independent spin-$s$ generators $T_{a_1...a_{s-1}}$ is precisely $2(s-1)+1$, that is, the dimension of a spin-$s$ (or, rather, $s-1$) representation of $\mathrm{sl}(2|\mathbb{R})$. The nice thing about it is that, because of the isomorphism $\mathrm{sl}(2|\mathbb{R}) \simeq \mathrm{so}(2,1)$, the components of our Chern–Simons connection corresponding to the spin-$s$ field come in the right number to form an irreducible spin-$(s-1)$ representation of the three-dimensional Lorentz group, $\mathrm{so}(2,1)$. Actually, this is exactly what we shall assume (and justify later on), namely that the spin-$s$ generators behave as irreducible Lorentz tensors, which can be seen to translate to

$$[T_a, T_{a_1...a_{s-1}}] = \epsilon^c{}_{a(a_1} T_{a_2...a_{s-1})c}. \tag{2.20}$$

---

[10] Obviously the off-shell degrees of freedom are what we shall map to those described by the Chern–Simons connections, for we know that on-shell three-dimensional higher-spins propagate no degrees of freedom at all.

The higher-spin algebra we are looking for is therefore some algebra containing $\mathrm{sl}(2|\mathbb{R})$ and, besides, the higher-spin generators $T_{a_1...a_{s-1}}$ up to some spin, sitting in irreducible representations of the Lorentz algebra according to the above formula. Note that the mismatch between the spin of some generators and the representation of $\mathrm{so}(2,1)$ they sit in comes from the fact that, on top of the frame indices, the connection further carries a spacetime index. The generators $T_{a_1...a_{s_1}}$, that we have said to have spin-$s$, are also sometimes said to have conformal spin $s-1$.

Two important comments are now in order. Firstly, it should be noted that, whatever our higher-spin algebra is in the end, in order to make sense of the Chern–Simons term it should be equipped with an invariant and non-degenerate bilinear form. If the searched-for algebra is semi-simple, then we know that the Killing form, which always exists, is non-degenerate (Cartan's Criterion). Moreover, if the algebra is simple, the Killing form is unique in the space of invariant bilinear forms. Interestingly, one can check that the only possibility for a bilinear form is[11]

$$(\Gamma, \Gamma) = \sum_{s=1}^{N} c_s \Gamma^{a_1...a_{s-1}} \Gamma_{a_1...a_{s-1}}, \tag{2.21}$$

where the coefficients $c_s$ are left undetermined by the requirement of invariance under the commutation relations we already have at hand, namely those of $\mathrm{sl}(2|\mathbb{R})$ as well as those in (2.20). The commutation relations among the higher-spin generators can potentially fix (some of) those coefficients, but for semi-simple Lie algebras there is at least one form (the Killing form) which corresponds to all the above coefficients being non-zero.

**Remark** : in the following we study finite- and infinite-dimensional algebras. Let us note, then, that in the infinite-dimensional case the definition of being simple is less clear. At any rate, the point is really to be able to construct a bilinear form with the desired properties, which we shall do anyhow, even for infinite dimension (see below).

The second comment to be made is that, assuming all $c_s$'s to be non-zero, whatever algebra we find will do the job. Namely, if we write a Chern–Simons theory for a gauge connection living in some higher-spin algebra containing $\mathrm{sl}(2|\mathbb{R})$ and whose higher-spin generators satisfy (2.20), the linearization thereof shall yield precisely the action (2.18) (upon identification of the components along the lines of $A = e + \omega$, and correspondingly for the second copy). This last point is really the key one, so let us phrase it differently: *when one linearizes the Chern–Simons action with proper identification of the degrees of freedom as above, the commutator of higher-spin generators with themselves is not used.* Only the commutators of higher-spin generators with spin-2 ones, and of course those of spin-2 generators with themselves are used. The reason for this is simple and lies in the fact that the higher-spin dreibeins and spin-connections have been assumed to have zero background values, as is easy to note trying to do the exercise. Another nice feature is that the coefficients $c_s$ are not used either when linearizing the action; indeed, at the free level the Chern–Simons term for our higher-spin algebra splits into a sum of free actions for the different spins which are involved, with the corresponding $c_s$ coefficients in front, which therefore play no role in recovering the Fronsdal system.

---

[11]Note that the unicity of such a form assumes, implicitly, that its formulation is covariant.

**Remark** : to be precise, it is the absolute value of the $c_s$ coefficients which plays no role in recovering the Fronsdal system. The relative signs of the coefficients are of some importance. Indeed, if the relative sign for the spin-2 and spin-3 sector is minus then the kinetic terms of both those sectors will have opposite signs, which is non-standard. This means that our above statement about the fact that the 'higher-higher' commutators do not affect the linearized limit of the theory should be refined: those commutators may constrain the bilinear form, which in turn may yield non-standard relative signs for the kinetic terms (if it is not degenerate). The example of the two non-compact real forms of sl(3), treated below, illustrates this point well.

The conclusion is thus that any Lie algebra containing sl(2|$\mathbb{R}$) whose higher-spin generators are irreducible Lorentz tensors and whose invariant bilinear form is non-degenerate shall yield a Chern–Simons action (with proper identification of the degrees of freedom) which, at the linearized level, agrees with the aforegiven free higher-spin system. The beauty of it is that we have reduced the quest for an interacting higher-spin theory in three dimensions to an algebraic problem: that of finding some Lie algebra satisfying the above requirements.

Two points now deserve a clear stating. The first is about simplicity and the second is about diversity, and we shall expand on them in the following. The 'simplicity' aspect is that something as common and easy to deal with as sl($n$|$\mathbb{R}$) fits into the above scheme. The 'diversity' aspect is that many other Lie algebras satisfy the requirements. We shall now proceed to expanding on those two points. However, let us already point out that we shall primarily be interested in infinite-dimensional algebras, so that the following part on finite-dimensional algebras is included for the sake of generality and so that one can compare it to the treatment of infinite-dimensional ones, addressed afterward.

### 2.2.2 Finite dimensional algebras

To the reader unfamiliar with the subject it might now come as a (good) surprise that, as we just said, something as "simple" as sl($n$) fits in this scheme [17]. Its usual presentation is the set of $n \times n$ traceless matrices (which is really the $n$-dimensional representation of it), but there exists another presentation. Indeed, consider the $n$-dimensional representation of sl(2|$\mathbb{R}$) and define the higher-spin generators to be the symmetrized products of the corresponding number of spin-2 generators (in their $n$-dimensional representation) minus the corresponding trace projections. One can then prove that the resulting algebra is in fact sl($n$), where $n-1$ is the maximum number of spin-2 generators we allow ourselves to take products of. As an example we give the commutation relations of sl(3) in this way:

$$
\begin{aligned}
{[T_a, T_b]} &= \epsilon_{abc} T^c, \\
{[T_a, T_{bc}]} &= \epsilon^m_{a(b} T_{c)m}, \\
{[T_{ab}, T_{cd}]} &= \sigma(\eta_{a(c}\epsilon_{d)bm} + \eta_{b(c}\epsilon_{d)am}) T^m,
\end{aligned}
\tag{2.22}
$$

where the $T_a$'s are defined to be the sl(2) generators in their three-dimensional representation and the $T_{ab}$'s are defined as

$$
T_{ab} \equiv T_{(a} T_{b)} - \frac{1}{3} \eta_{ab} T_c T_d \eta^{cd} = T_{ba}, \tag{2.23}
$$

a definition implying not only that $\eta^{ab} T_{ab} = 0$ identically but also that the $T_{ab}$'s themselves are traceless matrices (as can be checked), so that we are indeed reproducing some Lie algebra of traceless matrices, as is sl(3).

Note the presence of the $\sigma$ parameter in the commutator of two spin-3 generators, which labels the real form which is chosen. In fact, its absolute value can be changed by rescaling the generators, but its sign cannot; $\sigma < 0$ corresponds to sl(3|$\mathbb{R}$) while $\sigma > 0$ corresponds to su(1,2), the other non-compact real form of sl(3). As we have already pointed out, the last commutator hereabove does not affect the linearized limit, except for the relative sign of the spin-2 and spin-3 kinetic terms, with $\sigma < 0$ yielding a non-standard minus sign. Apart from those considerations (see below), any real form is thus a priori acceptable. Also, as can be checked, the bilinear form (2.21) is in this case non-degenerate, because sl(3) is simple.

The above scheme of things actually extends to the arbitrary-$n$ case of sl($n$), of which any non-compact real form is suited (a priori) to describe an interacting theory of higher spins up to spin $n$. The most used form, however, is sl($n$|$\mathbb{R}$), which is indeed very much studied in the literature. The reason for this is partly that it is simple to handle, and partly that for other real forms some of the kinetic terms for different higher-spins would have opposite relative signs.

**Remark** : of course since no on-shell degrees of freedom are propagated by our three-dimensional action one might wonder how relevant is the requirement that different higher-spin kinetic terms have the same relative sign (which is usually required to preserve unitarity). However, other pathological features may be seen to show up when using those different real forms, such as non-unitarity of the associated boundary theory [22].

What about other Lie algebras ? As is well known, any non-compact simple algebra contains sl(2|$\mathbb{R}$) as a subalgebra and, moreover, all semi-simple Lie algebras admit non-compact real forms. However, one might still wonder about the spectrum, that is, the requirement of containing, besides sl(2|$\mathbb{R}$), higher-spin generators forming irreducible representations of the three-dimensional Lorentz group. Actually, this is *also* guaranteed !

The argument is the following: consider any Lie algebra containing sl(2|$\mathbb{R}$) as well as other generators, that we collectively denote $T_A$. Assuming that our algebra is of finite dimension, the generators $T_A$ form a direct sum of finite-dimensional representations of sl(2|$\mathbb{R}$). The reason for it is the following: all of the $T_A$'s, taken together, certainly form some (finite-dimensional) representation of sl(2|$\mathbb{R}$) (which can be seen by considering the matrices corresponding to the sl(2|$\mathbb{R}$)-generators in the adjoint representation). Then, either this representation is irreducible, in which case we are done, or it is not, in which case it will split in some direct sum of irreducible representations (because of Weyl's theorem stating that any finite-dimensional representation of a semi-simple Lie algebra is completely reducible [23]).

The outcome of this analysis is thus that *any* non-compact form of *any* simple Lie algebra beyond sl(2|$\mathbb{R}$) is suited to describe some higher-spin theory via the Chern–Simons picture. Of course, and this is an important precision, some of them might actually contain higher-spin generators for only *some* spins beyond spin 2, that is, the spectrum might not be that of one irreducible representation of every spin up to some value. Furthermore, the spectrum might even contain spins below spin-2, and moreover the kinetic terms may in general enter the action with some relative signs (see below).

As a final comment before moving on to the infinite-dimensional higher-spin algebras we shall point out that, given some non-compact algebra, in general one may declare different sets of (three) generators to be the sl(2|$\mathbb{R}$) subalgebra describing pure gravity. Making such a choice is called choosing some 'embedding' of sl(2|$\mathbb{R}$) into the higher-spin algebra. Among all possible embeddings, there is a special one, called 'principal embedding', which has the property that all the other generators split into irreducible representations with multiplicity one.[12] Differently put, it means that the rest of the generators should organize as (2.20), once for each spin present in the spectrum. In the case of sl(3), for example, there is only one non-principal embedding, corresponding to the splitting of sl(3) as $\mathbf{8} = \mathbf{3} \oplus 2 \times \mathbf{2} \oplus \mathbf{1}$, whereas the principal embedding that we have presented in (2.22) corresponds to the splitting $\mathbf{8} = \mathbf{3} \oplus \mathbf{5}$ (the representations are denoted in boldface by their dimensions). Let us note that non-principally embedded sl(2|$\mathbb{R}$)'s have also been studied and seem somewhat more difficult to analyze. In particular, the properties of the corresponding boundary theory seem to present some subtleties — see e.g. [25, 26]. In the sequel, when we study infinite-dimensional higher-spin algebras, the gravitational sl(2|$\mathbb{R}$) subalgebra shall be principally embedded therein.

Note that all sl($n$) Lie algebras admit a non-compact form such that sl(2|$\mathbb{R}$) can be principally embedded thereof. Actually, some embeddings of sl(2|$\mathbb{R}$) may not be compatible with some non-compact real forms of whatever higher-spin algebra we use. For example, we point out that for the case of sl($n$) the principal embedding thereof is only compatible with the maximally non-compact real form, sl($n|\mathbb{R}$), as well as with su($\frac{n}{2}, \frac{n}{2}$) (or su($\frac{n-1}{2}, \frac{n+1}{2}$) if $n$ is odd). Let us also mention that the maximally non-compact real form is compatible

---

[12]An equivalent definition [24] is that the number of irreducible representations appearing in the spectrum is smaller than the rank of the algebra (which is $n$ for sl($n$)).

with any embedding and, conversely, the so-called 'normal' embedding is compatible with any real form. Last of all we also point out that one switches non-compact real forms for the principal embedding by multiplying all odd-spin generators by a factor of $i$, and for more information on such algebraic aspects we refer to [23].

### 2.2.3 Infinite-dimensional algebras

In the previous subsection we have been concerned with finding some completion to the commutation relations of sl$(2|\mathbb{R})$ together with (2.20). However, explicitly or implicitly, so far we have confined ourselves to exploring finite-dimensional Lie algebras. In the present subsection we address the question of infinite-dimensional higher-spin algebras. However, as their study is beyond the scope of the present notes we shall limit ourselves to making some comments on the subject.

The idea is that, along the lines of the construction of the sl(3) higher-spin generators in terms of products of spin-2 ones (see above), we may very well consider the same construction without limiting the degree of the products thereof. In such a way one would generate an infinite tower of higher-spin generators in representations of sl$(2|\mathbb{R})$. Such a construction of an infinite-dimensional (associative) algebra is actually rather standard and bears the fancy name of *universal enveloping algebra*, and it is denoted by $\mathcal{U}(\text{sl}(2|\mathbb{R}))$. Moreover, the universal enveloping algebra is some abstract construction [27] in which we build the higher-spin generators as products of the original ones for some abstract associative product, without considering the latter to be in some representation.[13] This is why, before obtaining our infinite-dimensional higher-spin algebra out of such a construction, there is one last step we need to perform; namely, quotienting by some value of the sl$(2|\mathbb{R})$-Casimir $C_2 \equiv T_a T_b \eta^{ab}$. The Lie algebra we are looking at is thus

$$\text{B}[\lambda] \equiv \text{hs}[\lambda] \oplus \mathbb{I} \equiv \frac{\mathcal{U}(\text{sl}(2|\mathbb{R}))}{\langle C_2 - \frac{1}{4}(\lambda^2 - 1)\mathbb{I} \rangle}, \tag{2.24}$$

where hs$[\lambda]$ is the standard infinite-dimensional higher-spin algebra in three dimensions, first introduced in [28] and then firstly explored by the authors of [29, 30, 31, 32, 33, 34]. Note that in the above expression we have also removed the identity, which is strictly speaking included in the universal enveloping construction, but which forms an ideal we are not interested in. Also note that, at this point, as defined by the above equation hs$[\lambda]$ is only an associative algebra, because so is $\mathcal{U}(\text{sl}(2|\mathbb{R}))$. However, in the sequel we shall equip it with the natural bracket (the antisymmetrization of the associative product) so to make it a Lie algebra, and we shall keep the same notation hs$[\lambda]$, which indeed usually denotes the Lie structure.

Quotienting as in the above relation is precisely the equivalent of considering the original sl$(2|\mathbb{R})$ generators to be in some representation, in which $C_2$ thus has *some* value $\lambda$,[14] and then taking products thereof. The parameter $\lambda$ is usually referred to as the *deformation parameter*, for reasons that shall become clear in the following. Let us further note that, in a theory of infinitely many scalar-coupled higher-spin fields in dimension three, the deformation parameter is related to the mass of the scalar [35, 36].

---

[13] The universal enveloping techniques can be applied to any Lie algebra [27].
[14] The $\lambda$ parameter is also sometimes denoted by $\mu$.

Differently put, we also need to match the desired spectrum, namely that of the correct number of generators $(2s+1)$ at each spin-level $s$, hence the need for quotienting. Indeed, if one does not quotient, the algebra actually contains an infinite number of spin-$s$ generators for a given $s$. That is because, if one does not identify $C_2$ with some value, then the trace of a spin-$s$ generator will be something transforming as a spin-$(s-2)$ generator but independent of those built by taking products of $s-3$ spin-2 ones (and one can take further traces). By quotienting one precisely relates those two kinds of objects, and a non-degenerate spectrum is thus obtained. Yet another way to understand the need for quotienting is to note that, otherwise, the Casimir would generate an ideal and the scalar product thereon (to be defined below) would then be degenerate. Let us also refer to [37] for a treatment of the universal enveloping construction in $\text{AdS}_D$.

As is guaranteed by the universal enveloping technique, $\text{sl}(2|\mathbb{R})$ is a subalgebra of $\text{hs}[\lambda]$, just as in the case of $\text{sl}(3)$ described above. Moreover, the algebra $\text{hs}[\lambda]$ contains higher-spin generators in irreducible representations of $\text{sl}(2|\mathbb{R})$: thanks to the quotienting by some value of the Casimir, there are $2(s-1)+1$ spin-$s$ generators for each $s = 2, 3, \ldots$ — the $\text{sl}(2|\mathbb{R})$ generators being understood as having spin 2, as before. In this way we thus manage to build suitable higher-spin algebras (up to the existence of appropriate bilinear forms thereon), and one might even think about universally enveloping other Lie algebras containing $\text{sl}(2|\mathbb{R})$, such as $\text{sl}(3|\mathbb{R})$. This would potentially yield higher-spin theories with different spectra. However, this approach has been largely ignored in the literature, with [38] among the exceptions (see also [37]). Higher-spin theories based on $\text{hs}[\lambda]$ are the most commonly studied among the infinite-dimensional ones.

**Remark** : it is of course wrong that any Lie structure comes from some associative one, and therefore on top of the aforementioned generalizations one could formally wonder about higher-spin Lie algebras whose Lie bracket is not the commutator of some associative product. Although in dimension four and greater it has been shown that such situations cannot arise [39], in dimension three, where at any rate one seems to have much more freedom, no such result has been obtained.

Besides infinite-dimensional subalgebras, one may of course wonder about the relation between $\text{hs}[\lambda]$ and its finite-dimensional cousins. However, it is an important point that $\text{sl}(n)$ is *not* a subalgebra thereof for $n \geq 3$. In fact, even for general $\lambda$, there is no finite-dimensional subalgebra of $\text{hs}[\lambda]$ apart from $\text{sl}(2|\mathbb{R})$ [32]. However, let us point out that for integer values $\lambda = N$, $\text{hs}[\lambda]$ does decompose into the sum of $\text{sl}(n|\mathbb{R})$ and an infinite-dimensional ideal one can then quotient by [32]. As we shall be working at $\lambda = \frac{1}{2}$ we do not dwell on that point any longer, and shall simply point out that such is the reason why $\text{hs}[\lambda]$ can be sometimes thought of as the 'analytic continuation' of $\text{sl}(n|\mathbb{R})$. Note that there are infinite-dimensional subalgebras, such as the well-known one consisting of only the even-spin generators (odd powers of the spin-2 ones), that one can restrict oneself to in a consistent fashion.

Let us now turn to describing a way of realizing hs[$\lambda$]. Indeed, our higher-spin algebra hs[$\lambda$] is compactly defined by the above universal enveloping expression, but the latter does not grant one with any convenient way of handling it. Of course, one can always work out the commutation relations among higher-spin generators from the above definition (along the lines of the sl(3|$\mathbb{R}$) case), but such an approach is far from handy. A far more convenient mean of treating our algebra is the so-called *oscillator realization*, which can be thought of as a refinement of the universal enveloping procedure (in the sense that it automatically imposes non-trivial conditions on the spectrum), and we now introduce it. The starting point is to notice that the spin-2 sector, sl(2|$\mathbb{R}$), can be realized in the following way. Consider a pair of commuting 'oscillators' $q_\alpha$, with $\alpha = 1, 2$, satisfying the following relation:

$$[q_\alpha, q_\beta]_\star \equiv 2i\epsilon_{\alpha\beta}, \tag{2.25}$$

where $\epsilon_{\alpha\beta} = -\epsilon_{\beta\alpha}$ is the two-dimensional $\epsilon$-symbol with conventions $\epsilon_{12} \equiv 1 = \epsilon^{12}$, with which we raise and lower the spinor indices of the $q_\alpha$'s. The $\star$-symbol denotes the associative product which the above Lie bracket corresponds to, and one can also formulate the above definition in terms of that product:

$$q_\alpha \star q_\beta \equiv q_\alpha q_\beta + i\epsilon_{\alpha\beta}, \tag{2.26}$$

which is called the $\star$-product [40, 41]. One then defines the quadratic combinations to be

$$T_{\alpha\beta} \equiv -\tfrac{i}{4} q_{(\alpha} \star q_{\beta)} = -\tfrac{i}{4} q_\alpha q_\beta, \tag{2.27}$$

where only three of them are independent, because $T_{12} = T_{21}$. The key observation is now that the Lie algebra of the quadratic generators $T_{\alpha\beta}$ is precisely sl(2|$\mathbb{R}$). Indeed, with the above definitions one easily checks that

$$[T_{\alpha\beta}, T_{\mu\nu}]_\star = -\tfrac{1}{2}(\epsilon_{\nu\beta} T_{\mu\alpha} + \epsilon_{\mu\beta} T_{\nu\alpha} + \epsilon_{\nu\alpha} T_{\mu\beta} + \epsilon_{\mu\alpha} T_{\nu\beta}), \tag{2.28}$$

which can be seen to reproduce the familiar sl(2|$\mathbb{R}$) commutation relations upon performing redefinitions.

Evidently, such a realization does not make the handling of sl(2|$\mathbb{R}$) any simpler — quite the contrary — but it allows us to generalize it in the following, natural way. Let us no longer restrict our attention to quadratic combinations of our oscillators and allow instead for generators of arbitrary degree (higher than two) in the $q$'s. Having in mind the universal enveloping construction, we shall nonetheless restrict the degree of the generators to be even (and the identity component is not included). In this way one generates an infinite dimensional Lie algebra, and one can check that it corresponds to hs[$\lambda$] for some value of $\lambda$. Indeed, the higher-spin generators defined as symmetric products of the oscillators (of even degree) do correspond to taking symmetrized products of our spin-2 generators, so that the construction is really a reformulation of the universal enveloping technique. Moreover one can verify that the above higher-spin generators do form irreducible representations of the spin-2 sector. However, one may wonder where is the deformation parameter, $\lambda$, in such an oscillator construction. In fact, in the latter realization the quotient is automatically

taken! The reason why this can happen is because we have specified more than the product (or commutation) relations among the $T_{\alpha\beta}$'s: we also know about how the $q_\alpha$ oscillators themselves commute to each other. To be fully convinced we should compute the Casimir $C_2$ in this formulation, the comparison of which with the universal enveloping construction tells us that the above oscillator realization corresponds to $\lambda = \frac{1}{2}$. Such a value of the deformation parameter is called the *undeformed case*, for reasons that shall me made clear hereafter. Thus, we can realize our higher-spin algebra at $\lambda = \frac{1}{2}$ as the algebra of linear combinations of our generators $T_{\alpha_1...\alpha_s}$ of even degree $s$ in the $q_\alpha$'s under the Lie bracket derived from (2.25).

At this point we owe it to the reader to answer the following question: what about hs$[\lambda]$ at $\lambda \neq \frac{1}{2}$? The answer is positive: there is a way to *deform* the oscillator relations (2.25) in such a way as to generate, upon considering generators of arbitrary (even) degree, hs$[\lambda]$ at generic $\lambda$ [32]. The so-called *deformed* oscillator relations read

$$[q_\alpha, q_\beta]_\star \equiv 2i(1 + (2\lambda - 1)K)\epsilon_{\alpha\beta}, \tag{2.29}$$

where $K \equiv (-)^{N_q}$ is the so-called Klein operator and $N_q$ counts the number of $q$ oscillators to its right, so that $K$ behaves as $(-)^{N_q} q_\alpha = -q_\alpha(-)^{N_q}$. One can check that the quadratic sector still reproduces sl$(2|\mathbb{R})$ and is independent of $\lambda$, but the higher commutation relations will of course depend on the deformation parameter. One might ask, however, whether the hs$[\lambda]$ algebras at different values of $\lambda$ are really different algebras or whether they are isomorphic, and it was shown that they differ [29, 30]. Evidently, at $\lambda = \frac{1}{2}$ one recovers the undeformed commutation relations of (2.25).

We have thus managed to realize a set of generators of the hs$[\lambda]$ algebra as powers of our deformed oscillators $q_\alpha$ (we keep the same notation). However, it might be felt that the above approach is not making the handling of hs$[\lambda]$ particularly simple. Indeed, the procedure to compute the $\star$-product (or $\star$-commutator) of two higher-spin generators involves successively making use of the formula (2.26) and identifying the produced generators. The structure of the $\star$-product of two higher-spin generators is clear but the details have to be worked out in quite a painful manner. As it turns out, there is a simpler way to deal with the oscillators, but which only works for the undeformed case: instead of taking our generators to be defined as symmetric $\star$-products of the oscillators we define them as simple products of the $q_\alpha$'s, and we further define the $\star$-product of any two polynomials $f$ and $g$ in the $q_\alpha$'s to be

$$(f \star g)(q'') \equiv \exp\left(i\epsilon_{\alpha\beta} \frac{\partial}{\partial q_\alpha} \frac{\partial}{\partial q'_\beta}\right) f(q)g(q')\big|_{q=q'=q''}, \tag{2.30}$$

where $f(q) \equiv f(q_1, q_2)$ and so on. In this way we have 'solved' for the $\star$-product, and one can check the above formula to imply the relations (2.25). Moreover, we define the Lie bracket defining our Lie algebra of $q$-polynomials to be

$$[f, g]_\star \equiv \frac{1}{2i}(f \star g - g \star f), \tag{2.31}$$

where the prefactor is a matter of conventions. It might seem as if the definition of the $\star$-product is now more complicated (it involves the exponential of differential operators), but it is really simpler, in the sense that we now have an explicit expression for it. Moreover, the above product law is in fact the so-called Moyal bracket, more familiar in the context of Quantum Mechanics. Again, one checks that the relations (2.25) and (2.26) are reproduced, so that the algebra is indeed the same.

Our algebra is thus simply the space of polynomials of even degree (but zero) in the oscillators $q_\alpha$, equipped with the above Lie bracket. A downside of the latter standpoint is that it cannot be generalized to arbitrary $\lambda$, namely, an explicit expression such as (2.30) cannot be obtained in the deformed case.[15] To the best of our knowledge, this curious fact still lacks a deeper justification, if there is any.

As a closing remark we simply mention that supersymmetric extensions exist, and we refer for example to [42, 43] in addition to [6] and [44]. More generally, for more information on the subject and in particular on supersymmetrizations thereof we refer to [45], where the reader shall also find additional discussions on real forms and reality conditions on these algebraic structures.

## 3. Comments on recent developments

The goal of these lectures has now been reached; namely, to formulate and explore the space of non-linear higher-spin theories in three dimensions. However, it would be frustrating to stop here without at least briefly commenting on what one can actually do with the theories we have built. We thus propose hereafter a very short and *non-exhaustive* review of recent developments referring to a *non-exhaustive* bibliography. The focus, of course, shall be on three-dimensional aspects.

**Coupling to matter** It should be noted that in the present text the construction of interacting higher-spin theories has been confined to the gauge sector, and couplings to matter have not been discussed. However, scalar couplings therewith can and were achieved by Prokushkin and Vasiliev in [46], where gauge invariance was shown to force the mass to take some value depending on the deformation parameter $\mu$ of the gauge algebra. However, that work was carried out at the level of the equations of motion, and what the corresponding action term should be is not clear at all. At the cubic level it is of course rather straightforward to build spin $0-0-s$ couplings by the usual method of simply contracting higher-spin currents (scalar bilinears involving derivatives with free indices) with higher-spin gauge fields. However, it is still an open issue how to go beyond that order.

**Metric formulation** As explained in detail in the present notes, the interactions have been introduced making use of the Chern–Simons picture. One might thus wonder whether it is possible to translate the full action (or EoMs) back to the metric-like formalism. However, as it turns out, such a task is highly non-trivial and there are ambiguities in how one writes the higher-spin fields (in metric-like formalism) in terms of the corresponding frame fields. In [16] the spin-3 was analyzed but it is not yet known how one can do it unambiguously for any spin, although a proposal does exist up to spin 5 [47]. One might of course choose not to "care" about the metric-like formalism and declare the frame (or gauge) picture to be more fundamental, but one is then faced with interpretation problems, for example when considering black-hole solutions with higher-spin charges turned on (see next point).

---

[15] There is a way to achieve this but one needs use a different set of oscillators [38].

**Black-Hole solutions** Just as for general relativity in three dimensions, which admits the famous BTZ black-hole solution [48, 49] when a negative cosmological constant is turned on, we expect the study of black holes in higher-spin gravity to be just as enlightening, if not more. And indeed the results obtained so far indicate a wild enrichment of the space of black-hole solutions as well as conceptually challenging thermodynamical issues, both of which deserve more studying. Among many references we point out the review [50].

**Quantization** The question of quantizing the theory, although very natural and interesting, seems to be difficult and has not been pursued much. However, one can always understand the holographic correspondence with some two-dimensional conformal field theory as a way of quantizing the theory (see next point). On the subtle and quite unexplored issue of quantizing in the standard way we shall not comment more.

**AdS$_3$/CFT$_2$** This question is one of the most important ones within the three-dimensional higher-spin context. Its study was essentially triggered in 2010 by [17, 51], where the asymptotic symmetry algebra of some three-dimensional higher-spin theories were computed and showed to correspond to some *W*-algebras. Now, the so-called *W*-algebras were known since the eighties, when their study started with the pioneering work of Zamolodchikov [52] and continued quite intensively in the following decade — see [53] for a review — , and they were known to be the global symmetries of some of the simplest conformal field theories in two dimensions, namely the minimal models [53], which later on would be seen to be the ones entering the correspondence. Moreover in this "low-dimensional" realization of the holographic principle one hopes to have the correspondence under more control than in the more standard framework (dimensions four and higher in the bulk) and yet, because of the higher spins present in the bulk, the problem should be non-trivial and possibly retain some of the interesting features of holography. The interest in this "non-trivial yet more tractable" realization of AdS/CFT thus grew and the past few years have witnessed considerable progress therein, with the notable contribution of Gaberdiel and Gopakumar, who first attempted at a comprehensive description of the correspondence [54]. Many successful checks thereof have since then been done but still some subtle issues remain, such as the presence of so-called light states in the bulk. We recommend [55] for a recent review of this most promising field.

**Acknowledgments**

It is a pleasure to thank A. Campoleoni and E. Skvortsov for their help in the understanding of some of the matters developed in the present work as well as for proofreading these notes, although the author should be held responsible for any mistakes to be found therein. We are also grateful to M. Moskovic for help in the preparation of the preprint. The author is also grateful for many interesting questions and comments by various people during the eighth edition of the Modave Summer School in Mathematical Physics. We also thank A. Rovai for communicating typos after publication, as well as P. Kessel for comments. We thank the Galileo Galilei Institute for Theoretical Physics for hospitality during the completion of this work. The work of G. L.G. is partially supported by IISN - Belgium (conventions 4.4511.06 and 4.4514.08), by the "Communauté Française de Belgique" through the ARC program and by the ERC through the "SyDuGraM" Advanced Grant. G. L.G. is a Research Fellow of the Fonds pour la Formation à la Recherche dans l'Industrie et dans l'Agriculture (F.R.I.A.).

## Appendix: the sl(2|ℝ) algebra

The sl(2|ℝ) Lie algebra can be realized as the real vector space of $2 \times 2$ traceless real matrices equipped with the usual Lie bracket

$$[M, M'] \equiv MM' - M'M, \tag{1}$$

where the multiplication is the matrix multiplication. Such matrices have the form

$$\begin{pmatrix} a & b \\ c & -a \end{pmatrix} \tag{2}$$

with $a, b, c$ real. For the standard (Chevalley–Serre) generators

$$H \equiv \begin{pmatrix} 1 & 0 \\ 0 & -1 \end{pmatrix}, \qquad E \equiv \begin{pmatrix} 0 & 1 \\ 0 & 0 \end{pmatrix}, \qquad F \equiv \begin{pmatrix} 0 & 0 \\ 1 & 0 \end{pmatrix} \tag{3}$$

we find the commutators

$$[H, E] = 2E, \qquad [H, F] = -2F, \qquad [E, F] = H, \tag{4}$$

with the usual matrix trace defining a scalar product

$$(M, M') \equiv \text{tr}(MM'), \tag{5}$$

which yields the non-zero projections

$$(H, H) = 2, \qquad (E, F) = (F, E) = 1. \tag{6}$$

The contact with the "covariant" formulation is made by the redefinitions

$$E \equiv T_1 + T_2, \qquad F \equiv T_2 - T_1, \qquad H \equiv 2T_3 \tag{7}$$

for which the commutations relations can be packaged into

$$[T_a, T_b] = \epsilon_{abc} T^c, \tag{8}$$

where indices are raised and lowered with the metric $\eta^{ab} \equiv (-++)$ and we use the convention $\epsilon_{123} = 1$. The bilinear form (6) here reads

$$(T_a, T_b) = \frac{1}{2} \eta_{ab}. \tag{9}$$

We also point out that, if the generators are no longer restricted to be real, a two dimensional representation of sl(2|ℝ) is given by

$$T_1 \equiv \tfrac{1}{2} \begin{pmatrix} -i & 0 \\ 0 & i \end{pmatrix}, \qquad T_2 \equiv \tfrac{1}{2} \begin{pmatrix} 0 & -i \\ i & 0 \end{pmatrix}, \qquad T_3 \equiv \tfrac{1}{2} \begin{pmatrix} 0 & -1 \\ -1 & 0 \end{pmatrix}, \tag{10}$$

which is a more natural basis if one thinks of sl(2|ℝ) as su(1,1) (they are isomorphic).

This representation can be generalized to a (reducible) $D$-dimensional representation by moving the non-zero entries to the "corners" of the $D \times D$ matrices and leaving a $(D-2)\times(D-2)$ block in the middle. As can be checked, this defines the so-called "normal" embedding of sl$(2|\mathbb{R})$ into sl$(D)$, and it is indeed easy to check that it is compatible with any non-compact form su$(p,q)$ sl$(D)$ (on top of the maximally non-compact one sl$(D|\mathbb{R})$).

Another $D$-dimensional representation, this time defining the so-called "principal embedding" of sl$(2|\mathbb{R})$, is obtained by setting (defining $d \equiv \frac{D}{2}$):

$$T_1 \equiv \tfrac{1}{2}\begin{pmatrix} -i\mathbb{I}_{d\times d} & \mathbb{O}_{d\times d} \\ \mathbb{O}_{d\times d} & i\mathbb{I}_{d\times d} \end{pmatrix}, T_2 \equiv \tfrac{1}{2}\begin{pmatrix} \mathbb{O}_{d\times d} & -i\mathbb{I}_{d\times d} \\ i\mathbb{I}_{d\times d} & \mathbb{O}_{d\times d} \end{pmatrix}, T_3 \equiv \tfrac{1}{2}\begin{pmatrix} \mathbb{O}_{d\times d} & -\mathbb{I}_{d\times d} \\ -\mathbb{I}_{d\times d} & \mathbb{O}_{d\times d} \end{pmatrix}, \qquad (11)$$

and correspondingly for the odd-$D$ case. As can be checked, this embedding is only compatible with the non-compact form su$(d,d)$ of sl$(D)$ (in addition to the maximally non-compact one), as it should be.